  \documentstyle[psfig]{MN}

\def\phm{\phantom{$-$}}
\def\ls{{_<\atop^{\sim}}}
\def\gs{{_>\atop^{\sim}}}
\def\lg{{_>\atop^{_<}}}

\def\refitem{\par\parskip 0pt\noindent\hangindent 20pt}

\def\etal{{et\thinspace al.}\ }
\def\eg{{\rm e.g.}\ }
\def\ie{{\rm i.e.}\ }

\title[A Unifying View of the SED of Blazars] 
{A Unifying View of the Spectral Energy Distributions of Blazars}

\author[G. Fossati, L. Maraschi, A. Celotti, A.  Comastri, G. Ghisellini]
{G. Fossati$^1$, L. Maraschi$^2$, A. Celotti$^{1,3}$, A. Comastri$^4$ 
and G. Ghisellini$^2$\\
$^1$ S.I.S.S.A., via Beirut 4, I--34014, Trieste, Italy \\
$^2$ Osservatorio Astronomico di Brera, via Brera 28, I-20121, Milano, Italy\\
$^3$ Institute of Astronomy, Madingley Road, Cambridge CB3 0HA \\
$^4$ Osservatorio Astronomico di Bologna, via Zamboni 33, I-40126, Bologna, 
Italy\\}

\date{Received ***; in original form ***}

\begin{document}
\maketitle

\begin{abstract}  
We collect data at well sampled frequencies from the radio to the 
$\gamma$--ray range for the following three complete--samples of blazars: 
the Slew Survey and the 1 Jy samples of BL Lacs and
the 2 Jy sample of Flat Spectrum Radio--Loud Quasars (FSRQs).  
The fraction of objects detected in $\gamma$--rays (E $\gs 100$ MeV) is 
$\sim$ 17 \%, 26 \% and 40 \% in the three samples respectively.  
Except for the Slew Survey sample, $\gamma$--ray detected sources do not 
differ either from other sources in each sample, nor from all the $\gamma$-ray 
detected sources, in terms of the distributions of redshift, radio and X--ray 
luminosities and of the broad band spectral indices (radio to optical and 
radio to X--ray).

We compute average spectral energy distributions (SEDs) from
radio to $\gamma$--rays for each complete sample and for groups of blazars
binned according to radio luminosity, irrespective 
of the original classification as BL Lac or FSRQ.
  
The resulting SEDs show a remarkable continuity in that: 
i) the first peak occurs in different frequency ranges for different samples/ 
   luminosity classes, with most luminous sources peaking at
   lower frequencies; 
ii) the peak frequency of the $\gamma$--ray component
    correlates with the peak frequency of the lower energy one;
iii) the luminosity ratio between the high and low frequency components
     increases with bolometric luminosity.

The continuity of properties among different classes of sources and the
systematic trends of the SEDs as a function of luminosity favor a unified
view of the blazar phenomenon: a single parameter, related to luminosity,
seems to govern the physical properties and radiation mechanisms in the
relativistic jets present in BL Lac objects as well as in FSRQ. 
The general implications of this unified scheme are discussed while a detailed
theoretical analysis, based on fitting continuum models to the individual 
spectra of most $\gamma$-ray blazars, is presented in a separate paper
(Ghisellini et al. 1998).

\end{abstract} 

\begin{keywords} quasars: general -- BL Lacertae objects: general --
X--rays: galaxies -- X--rays: general -- radiative mechanisms:
non--thermal -- surveys
\end{keywords}

\section{Introduction}

The discovery of BL Lac objects and the paradoxes associated with
their violent variability led to a major step forward in the theory of
Active Galactic Nuclei (AGN), that is to the concept of relativistic
jets. Flat spectrum, radio--loud quasars (Angel \& Stockman 1980)
share basically all of the properties of BL Lac objects related to the
presence of a strong non--thermal broad band continuum, except for
the absence of broad emission lines. Hence the common designation of
blazars proposed by Ed Spiegel in 1978.

It was initially supposed that BL Lacs represented the most extreme
version of FSRQs, i.e. those with the most highly boosted continuum.
Instead, it has been recognized later (e.g. Ghisellini, Madau \&
Persic 1987; Padovani 1992a; Ghisellini et al. 1993) that the {\it
amount of relativistic beaming} and the {\it intrinsic} power in the
lines are lower in BL Lacs than in FSRQs, implying some intrinsic
difference between the two classes. Differences are also found in the
extended radio emission and jet structure (e.g. Padovani 1992a;
Gabuzda et al. 1992). Nevertheless the continuity of several observational
properties including the luminosity functions (Maraschi \& Rovetti
1994), the radio to X--ray SEDs (Sambruna et al. 1996)
 and the luminosity of the lines (Scarpa \& Falomo 1997) suggests
that blazars can still be considered as a single family, where the
physical processes are essentially similar allowing for some scaling
factor(s).  The identification of these scaling factors would represent a
substantial progress in the understanding of the blazar phenomenon.

A special class of BL Lacs was found from identification of  X--ray sources.
X--ray selected BL Lacs (XBL) differ from the classical radio--selected BL
Lacs (RBL) in a lesser degree of ``activity'' (including polarization),
in the radio to optical emission and in the relative intensity of
their X--ray and radio emission. This led to the suggestion that the
X--ray radiation was less beamed than the radio one and that XBLs were
observed at larger inclination to the jet axis (e.g. Urry \& Padovani
1995 for a review). 

Giommi \& Padovani (1994) quantified the differences in SEDs
between XBLs and RBLs, and Padovani \& Giommi (1995) introduced 
the distinction between 'High--energy peak BL Lacs' (HBL) 
and 'Low--energy peak BL Lacs' (LBL), for objects which emit
most of their synchrotron power at high (UV--soft--X) or low (far--IR,
near--IR) frequencies respectively.
Quantitatively a distinction can be done on the basis of the ratio
between radio and X-ray fluxes (see also \S3.2.2). 
We will use the broad band spectral index $\alpha_{\rm
RX}$\footnote{Hereinafter we define spectral indices as 
F$_\nu \propto \nu^{-\alpha}$. In broad band indices radio, optical and
X--ray fluxes are taken at 5 GHz, 5500 \AA\ and 1 keV, respectively.}
and call HBL and LBL objects having  $\alpha_{\rm RX} \ls 0.75$, $\gs
0.75$, respectively. 
Giommi and Padovani also proposed that HBL represent a small 
fraction of the BL Lac population
and are numerous in X--ray surveys only due to selection effects.
An alternative hypothesis (Fossati et al. 1997) relates 
the spectral properties to the source luminosity in such a way that
low luminosity objects (with high space density)
are HBLs while high luminosity objects (with low space density) are LBLs.

We will include here X--ray selected BL Lacs
together with ``classical'' BL Lacs in the blazar family, again
assuming that the basic physical processes by which the continuum is
produced are common to the whole family.
 
The detection by  EGRET, on board the Compton Gamma--Ray Observatory (CGRO),
of many  blazars at $\gamma$--ray energies (E $\gs 30$ MeV) revealed that
a substantial fraction and in some cases the 
bulk of their power is emitted in this very high energy band.
The $\gamma$--ray emission  is therefore of fundamental 
importance in the SED of blazars.
 
From the theoretical point of view the radio to UV continuum is
universally attributed to synchrotron emission from a relativistic
jet, while a flat inverse Compton component due to upscattering of the
low energy photons is expected to emerge at high energies as originally
discussed in Jones, O'Dell \& Stein (1974).
The latter process is therefore a plausible candidate to explain the
$\gamma$--ray emission. 
The soft photons to be upscattered could be either the synchrotron photons 
themselves (synchrotron self--Compton process, SSC,
e.g. Maraschi, Ghisellini \& Celotti 1992; Bloom \& Marscher 1993) or
photons produced by the disk and/or scattered /reprocessed in the
broad line region (Blandford 1993; Dermer \& Schlickeiser 1993;
Sikora, Begelman \& Rees 1994; Ghisellini \& Madau 1996). 
Understanding whether/how the $\gamma$--ray properties differ among subclasses 
is essential to assess the role of different mechanisms and to verify
whether the idea of blazars as a unitary class can be maintained.
 
Here we study the systematics of the SEDs of blazars using data from the
radio to the $\gamma$-ray band. We confirm and extend  
previous results of Maraschi et al. (1995) and Sambruna et al. (1996) by: 
i) {\it extending the SED to the $\gamma$-ray range}; 
ii) using a much larger complete sample of FSRQ; 
iii) using the richer and brighter sample of X--ray selected BL Lacs 
recently derived from the Slew Survey.
We use the available $\gamma$--ray data for each sample but also indirect 
information derived from the $\gamma$--ray detected (not complete) sample
discussed by Comastri et al. (1997). Since we find
that the continuity hypothesis among blazars holds we also consider a
merged ``global'' sample subdivided in luminosity bins irrespective of
the original classification of the objects.

The structure of the paper is as follows. 
In section 2 we describe how the data for the SEDs were collected and 
treated for each sample.  The $\gamma$--ray properties of different samples 
are also discussed.  
In section 3 we build average SEDs for the three sub--samples and for the
global sample subdivided according to luminosity and present our results.
These are discussed in section 4, while our conclusions are drawn in
section 5.

\section{The data}

\subsection{The samples}

We decided to consider the following three samples of blazars: the
Slew Survey Sample and the 1 Jy sample of BL Lac objects and the FSRQ
sample derived from the 2 Jy sample of Wall \& Peacock (1985),
motivated by the need of completeness, sufficient number of
objects and observational coverage at other frequencies, as detailed
below.

\subsubsection{BL Lacs, X--ray selected: the Slew survey sample}

The {\sl Einstein} Slew survey (Elvis \etal 1992) was derived from
data taken with the IPC while the telescope scanned the sky in between
different pointings.
It has limited sensitivity (flux limit of $\simeq 5 \times 10^{-12}$ 
erg cm$^{-2}$ s$^{-1}$ in the IPC band [0.3 $-$ 3.5 keV]), but covers a large
fraction of the sky ($\sim$ 36000 deg$^2$).  
In a restricted region of the sky 
Perlman \etal (1996a) selected a sample of 48 BL Lacs 
(40 HBL, 8 LBL) which can                        
be regarded as being practically complete.  
This is the largest available X--ray selected sample of BL Lacs. 
The redshift is known for 41 out of the 48 objects, and 8/48 have been 
detected at $\gamma$--ray energies, (6 with EGRET, E$\gs$ 100 MeV, 1 with 
Whipple, E$\gs$ 0.3 TeV, 1652+398, and 1 with both instruments, 1101+384).

\subsubsection{BL Lacs, radio selected: the 1 Jy sample}

This is the largest complete radio sample of BL Lacs compiled so far.
The complete 1 Jy BL Lac sample was derived from the catalog of
extragalactic sources with $F_{\rm 5GHz} \ge 1$ Jy (K\"uhr \etal 1981)
with additional requirements on radio flatness ($\alpha_{\rm R} \le 0.5$),
optical brightness ($m_V \le 20$) and the weakness of optical emission 
lines (EW$_{\lambda} \le 5$ \AA, evaluated in the source rest frame) 
(Stickel \etal 1991). 
This yielded 34 (2 HBL, 32 LBL) sources matching the criteria, 26 with a 
redshift determination and 4 with a lower limit on it (Stickel, Meisenheimer 
\& K\"uhr 1994). 
Out of these 34 objects, 9 have been detected at $\gamma$--ray energies
(8 with EGRET, plus 1 with Whipple, 1652+398).

\subsubsection{Flat Spectrum Quasars: Wall \& Peacock sample}

For  FSRQs we considered  the sample drawn 
by Padovani \& Urry (1992) from the ``2 Jy sample'' (Wall \& Peacock 1985), 
a complete flux--limited catalogue selected at 2.7 GHz, covering 9.81 sr, 
and including 233 sources with $F_{\rm 2.7GHz} > 2$ Jy, 
and $\alpha_{\rm R} \le 0.5$.
It consists of 50 sources with almost complete polarization data, of
which 20 are detected in $\gamma$--rays (all with EGRET).

\subsubsection{The Total Blazar sample}

Combining the three samples yields a total of 126 blazars 
(six of them are present in both the radio and X--ray selected
samples of BL Lacs), of which 33 detected in $\gamma$--rays. 
We will refer to them as the {\it total blazar sample}.

\subsection{Multi-frequency Data}

In view of building average  SEDs minimizing the bias introduced by 
incompleteness, we decided to focus on a few well covered frequencies, 
at which fluxes are available for most objects. 
In a separate paper (Ghisellini et al. 1998) we consider a sub--sample 
of $\gamma$--ray loud blazars with extensive coverage in frequency with the 
scope of carrying out detailed model fitting for each source.

We chose the following seven well sampled frequencies, that are
sufficient to give the basic information on the SED
shape from the radio to the X--ray band: radio at 5 GHz, millimeter at 230
GHz, far infrared (IRAS data) at 60 and 25 $\mu$m, near infrared (K band) 
at 2.2 $\mu$m, optical (V band) at 5500 \AA, and soft X--rays at 1 keV.
Data were collected from a careful search in the literature and extensive 
usage of the NASA Extragalactic Database (NED)\footnote{
The optical magnitudes have been de--reddened using values of A$_{\rm
V}$ derived from the A$_{\rm B}$ reported in the NED database according to 
the law A$_{\rm V} =$ A$_{\rm B}/1.324$ (Riecke \& Lebofski 1985)}.
In the radio and optical bands the coverage is complete for all the
objects in the three samples, while unfortunately for mm, far and near
IR and X--ray fluxes data for some sources are lacking (see Table~1).
The worse case is the far IR (25 $\mu$m) band where only 28/126 
objects have measured fluxes.

For each source, at each frequency {\it from radio to optical} we assigned
the average of all the fluxes found in literature. 
Given the large variability these averages were performed logarithmically 
(magnitudes).

In principle a suitable alternative to the averaging would be 
to consider in each band the maximum detected flux (see for
instance Dondi \& Ghisellini 1995).
On one hand this choice could be particularly meaningful in view of the
fact that in the $\gamma$--ray band, due to the limited sensitivity of
detectors, we are biased towards measuring the brightest states.  
On the other hand also this option is biased since the value of the
maximum flux is strongly dependent on the observational coverage and for
most of the objects we only have a few (sometimes a single) observations.  
Moreover the strength of this bias is 'band--dependent' and can thus
significantly affect the determination of the broad band spectral shape.  
As both choices present advantages and disadvantages, and since our goal
is a statistical analysis, we consider them equally good.
The 'averages' option has been preferred because it is likely to be more
robust with respect to the definition of radio--optical SED properties.

In Table~1 a summary of the collected broad band data is reported,
with the computed average flux values for each object. 


\subsubsection{X--ray data}

The knowledge of the X--ray properties is of special relevance because
in this band both the synchrotron and inverse Compton processes can
contribute to the emission. Since the first mechanism is expected to
produce a steep continuum in this band while the second one should
give rise to a flat component ($\alpha \le 1$, rising in a $\nu
F_{\nu}$ plot) the shape of the X--ray spectrum can give a fundamental
hint for disentangling the two components and inferring the respective
peak frequencies.  

We privileged the large and homogeneous {\it ROSAT} data base.  In fact, a
large fraction of the 126 sources (90/126) has been observed with the {\it
ROSAT} PSPC allowing to uniformly derive X--ray fluxes and in many cases,
that is for 73 targets of pointed observations, spectral shapes in the
0.1--2.4 keV range (Brunner et al. 1994; Lamer et al. 1996; Perlman et al.
1996b; Urry et al. 1996; Comastri et al. 1995, 1997; Sambruna 1997).
X--ray spectral indices were derived from the same observation and, when
available, we adopted the $\alpha_{\rm X}$ resulting from fits with 
neutral hydrogen column density N$_{\rm H}$ allowed to vary.  
Some of these 90 objects (17) have been only detected in the {\it ROSAT}
All Sky Survey (RASS) and fluxes are published by Brinkmann, Siebert \&
Boller (1994) and Brinkmann et al. (1995).  
Monochromatic fluxes (at 1 keV) for these sources have been derived from
the 0.1 -- 2.4 keV integrated flux adopting the average spectral index of
the sample to which they belong (see Table~4) and the value of the Galactic
column in the source direction (Elvis et al 1989; Dickey \& Lockman 1990;
Lockman \& Savage 1995; Murphy et al 1996).
When more than one observation was available we give the average flux.

Of the remaining 36 sources, 24 belong to the Slew survey sample and for them 
we used directly the {\sl Einstein} IPC flux from Perlman et al. (1996a).  
The fluxes at 2 keV listed by Perlman et al. (1996a) were converted to 1 keV 
using the average {\it ROSAT} spectral index of the Slew survey 
sample ($\langle \alpha_{\rm X}\rangle$ =1.40), derived from the 24 sources 
with a {\it ROSAT} measured value. 

For other 3 sources, without ROSAT data, we used an {\sl Einstein} IPC 
flux, bringing the total number of sources with measured X--ray flux
to 117/126.

\subsubsection{$\gamma$--ray data}

Within the three samples only a fraction of blazars were detected in
$\gamma$--rays, namely 9/34 in the 1 Jy sample, 8/48 in the Slew
sample, 20/50 in the FSRQ sample.  Four of these sources (0235+164,
0735+178, 0851+202 and 1652+398) are present in both the BL Lac
object samples, giving a net number of $\gamma$--ray detections of 33
out of 126 blazars.  
All but one of them have been observed by EGRET in the 30 MeV -- 30 GeV band.  
For 28/32 a $\gamma$--ray spectral index has been determined.  
One source, 1652+398 (Mkn 501), has only been detected at very high energies,
beyond 0.3 TeV by ground based Cherenkov telescopes (Whipple and HEGRA,
Weekes et al. 1996; Bradbury et al. 1997), while EGRET yielded only an
upper limit.
It is worth noticing that the detected fraction is significantly different 
between quasars and BL Lacs, being respectively $40 \pm 10.6$ \% and 
$17.1 \pm 5.1$ \% for XBLs and RBLs together. However for RBLs only
the fraction detected in $\gamma$--rays is $26.5 \pm 9.9$ \%, consistent with 
that of quasars while XBLs only yield $16.7 \pm 6.4$ \%.

\begin{table}
\setcounter{table}{1}
\caption{Basic data for the 30 $\gamma$--ray detected sources not
included in our samples : 
(1): IAU name; 
(2): redshift; 
(3): radio flux at 5 GHz; 
(4): $\gamma$--ray flux at 100 MeV; 
(5): EGRET spectral index}
\begin{tabular}{@{}ccccc}
\hline
(1) &(2) &(3) &(4) &(5)  \\
IAU name & z & F$_{\rm 5GHz}$ & F$_{\rm 100MeV}$ & $\alpha_\gamma$ \\
 & & (Jy) & (nJy) &  \\
\hline
&&&& \\
{ 0130$-$171 } & { 1.022 } & { 1.00 } & { 0.122 } & { ... } \\ 
{ 0202+149 }   & { 1.202 } & { 2.40 } & { 0.383 } & { 1.5  $\pm$  0.1 } \\ 
{ 0234+285 }   & { 1.213 } & { 2.36 } & { 0.296 } & { 1.7  $\pm$  0.3 } \\ 
{ 0446+112 }   & { 1.207 } & { 1.22 } & { 0.470 } & { 0.8  $\pm$  0.3 } \\ 
{ 0454$-$234 } & { 1.009 } & { 2.2 }  & { 0.143 } & { ... } \\ 
{ 0458$-$020 } & { 2.286 } & { 2.04 } & { 0.364 } & { ... } \\ 
{ 0506$-$612 } & { 1.093 } & { 2.1 }  & { 0.062 } & { ... } \\ 
{ 0521$-$365 } & { 0.055 } & { 9.7 }  & { 0.139 } & { 1.16  $\pm$  0.36 } \\ 
{ 0804+499 }   & { 1.433 } & { 2.05 } & { 0.322 } & { 1.72  $\pm$  0.38 } \\ 
{ 0805$-$077 } & { 1.837 } & { 1.04 } & { 0.404 } & { 1.4  $\pm$  0.6 } \\ 
{ 0827+243 }   & { 0.939 } & { 0.67 } & { 0.226 } & { 1.21  $\pm$  0.47 } \\ 
{ 0829+046 }   & { 0.18 }  & { 1.65 } & { 0.132 } & { ... } \\ 
{ 0917+449 }   & { 2.18 }  & { 1.03 } & { 0.075 } & { 0.98  $\pm$  0.25 } \\ 
{ 1156+295 }   & { 0.729 } & { 1.65 } & { 1.727 } & { 1.21  $\pm$  0.52 } \\ 
{ 1222+216 }   & { 0.435 } & { 0.81 } & { 0.278 } & { 1.50  $\pm$  0.21 } \\ 
{ 1229$-$021 } & { 1.045 } & { 1.1 }  & { 0.250 } & { 1.92  $\pm$  0.44 } \\ 
{ 1313$-$333 } & { 1.210 } & { 1.47 } & { 0.098 } & { 0.8  $\pm$  0.3 } \\ 
{ 1317+520 }   & { 1.060 } & { 0.66 } & { 0.079 } & { ... } \\ 
{ 1331+170 }   & { 2.084 } & { 0.713 }& { 0.091 } & { ... } \\ 
{ 1406$-$076 } & { 1.494 } & { 1.08 } & { 1.013 } & { 1.03  $\pm$  0.12 } \\ 
{ 1604+159 }   & { 0.357 } & { 0.50 } & { 0.260 } & { 0.99  $\pm$  0.50 } \\ 
{ 1606+106 }   & { 1.227 } & { 1.78 } & { 0.312 } & { 1.20  $\pm$  0.30 } \\ 
{ 1622$-$297 } & { 0.815 } & { 1.92 } & { 2.416 } & { 1.2  $\pm$  0.1 } \\ 
{ 1622$-$253 } & { 0.786 } & { 2.2 }  & { 0.336 } & { 1.3  $\pm$  0.2 } \\ 
{ 1730$-$130 } & { 0.902 } & { 6.9 }  & { 0.258 } & { 1.39  $\pm$  0.27 } \\ 
{ 1739+522 }   & { 1.375 } & { 1.98 } & { 0.236 } & { 1.23  $\pm$  0.38 } \\ 
{ 1933$-$400 } & { 0.966 } & { 1.48 } & { 0.158 } & { 1.4  $\pm$  0.2 } \\ 
{ 2032+107 }   & { 0.601 } & { 0.77 } & { 0.192 } & { 1.5  $\pm$  0.3 } \\ 
{ 2344+514 }   & { 0.044 } & { 0.215 }& { 0.8$^{a}$ } & { ... } \\ 
{ 2356+196 }   & { 1.066 } & { 0.70 } & { 0.311 } & { ... } \\ 
&&&& \\
\hline
\end{tabular}
{\raggedright\par 
{\bf References of Table 2:}\par\small
References for the data here reported are listed in the Notes to Table~5.\par
}
{\raggedright 
{\bf Notes to Table 2:}\par\small
\noindent$^{(a)}$ source detected only by Whipple. 
The given value is the integrated flux measured at E $>$ 300 GeV, 
in units of $10^{-11}$ photons cm$^{-2}$ s$^{-1}$.\par
}
\medskip
\end{table}

Many other blazars ($\sim$ 30) have been detected by EGRET, but do not
fall in our samples. 
One can consider the group of $\gamma$--ray detected objects as a sample 
in its own right, though not a complete one at present, since a significant 
fraction of the sky has been surveyed though not uniformly.  
This larger sample comprises 66 sources (Fichtel et al. 1994; 
von Montigny et al. 1995; Thompson et al. 1995, 1997; Mattox et al. 1997
and references therein), of which 60 have a measured redshift, 
and 48 an estimate of the spectral index.   
To this set we can add Mkn 501 (already included in both our BL Lac samples), 
and 2344+514, detected only by the Whipple telescope (Fegan 1996). 
We will use this additional information to discuss whether the $\gamma$--ray 
properties of our samples can be representative of the whole $\gamma$--ray
loud population and if so, to increase the statistics (see section 3.1).
We therefore collected basic data also on all of the 30 (29 EGRET plus
2344+514) $\gamma$--ray detected AGN/blazars not included in the complete 
samples.  They are reported in Table~2.

Given the large amount of observations and analysis of the same data
by different authors, for the selection of the flux and spectral index
we used the following criteria: i) spectral index and flux referring
to the same observation, ii) data corresponding to a single viewing
period, iii) if data were analyzed by different authors, the results of
the most recent analysis are preferred. 

$\gamma$--ray data are usually given in units of photons cm$^{-2}$
s$^{-1}$ above an energy threshold (e.g. for EGRET E $\gs$ 100 MeV).
We converted them to monochromatic fluxes at 100 MeV integrating a
power law in photons with the measured or assumed (the average) spectral index.

\begin{table}
\caption{spectral indices used for K--correction of monochromatic
fluxes: (1) spectral band; (2) Slew; (3) 1Jy; (4) FSRQ; (5) references. }
\begin{tabular}{@{}clcccc}
\hline
& band & Slew & 1 Jy  & FSRQ & refs. \\
& {\hfil(1)\hfil}  &  (2) &  (3)  & (4) & (5) \\
\hline
&&&&&\\
& radio          & 0.20 & $-$0.27\phm & $-$0.30\phm & 1 \\
& mm             & 0.32 &    0.32     &    0.48     & 2 \\
& IRAS           & 0.60 &    0.80     &    1.00     & 3 \\ 
& IR--opt        & 0.67 &    1.21     &    1.52     & 4 \\ 
& X--rays        & 1.40 &    1.25     &    0.83     & 5 \\
& $\gamma$--rays & 0.98 &    1.26     &    1.21     & 5 \\
&&&&&\\
\hline
\end{tabular}
{\raggedright\par 
{\bf References to Table 3:}\par\small
(1) Stickel et al. 1994; 
(2) Gear et al. 1994; 
(3) derived from IRAS data; 
(4) Bersanelli et al. 1992;
(5) this work, see Table~5.
}
\medskip
\end{table}

\subsubsection{Luminosities and K--correction}

\noindent
All fluxes were K-corrected and luminosities were computed with the
following choices:

\begin{itemize}

\item[a)] we considered lower limits on redshift (4 sources) as detections, 
while we assigned the average redshift of the sample to the few 
sources without any estimate 
(4 in the 1 Jy sample, for which $\langle z \rangle = 0.492$, 
and 6 in the Slew survey sample, $\langle z \rangle = 0.194$);  

\item[b)] luminosity distances were calculated adopting H$_0$=50 km
s$^{-1}$ Mpc$^{-1}$ and q$_0$=0;

\item[c)] fluxes were K--corrected according to the following
prescriptions.  For radio--to--optical data we used average spectral
indices derived from the literature (see Table~3).  
For X--ray and $\gamma$--ray data we used measured power law spectral indices, 
when available, or the average index derived for the sources of the same 
sample (see Table~3).

\end{itemize}

\section{Results}

\subsection{Distributions of properties}

Since the fraction of objects detected in $\gamma$-rays in the three
samples is rather small, it is important to ask whether the detected
sources are representative of each sample as a whole or are
distinguished in  other properties from the rest of the objects in it.
Moreover we want to verify whether the $\gamma$-ray detected sources in
general differ from those belonging to the complete samples.
 
We therefore computed the distributions of various quantities, i.e. redshift, 
luminosities and broad band spectral indices, for objects belonging to 
each sample. 
These are shown in Figs.~1--5 for the three samples and the total blazar one, 
evidentiating those sources detected in the $\gamma$--ray band as grey shaded 
areas in the histograms.  
The redshift, radio (at 5GHz) and X--ray (at 1keV) 
luminosities (expressed as $\nu$L$_\nu$, erg/sec) are shown in Figs.~1--3, 
while the distributions of the broad band spectral indices $\alpha_{\rm RO}$ 
and $\alpha_{\rm RX}$ are plotted in Figs.~4 and 5, respectively.

\begin{figure}
\psfig{file=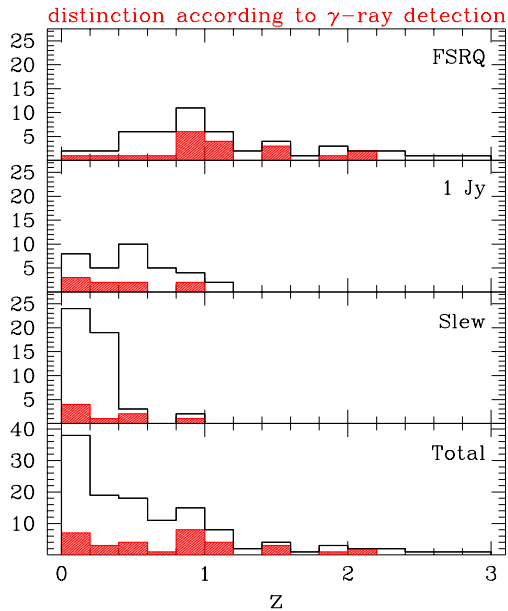,width=8.5truecm,height=8.5truecm}
\caption[h]{Redshift distributions for the three complete samples and
the ``total blazars sample". Sources detected in the $\gamma$--rays are
indicated by the grey areas.}
\end{figure}

\begin{figure}
\psfig{file=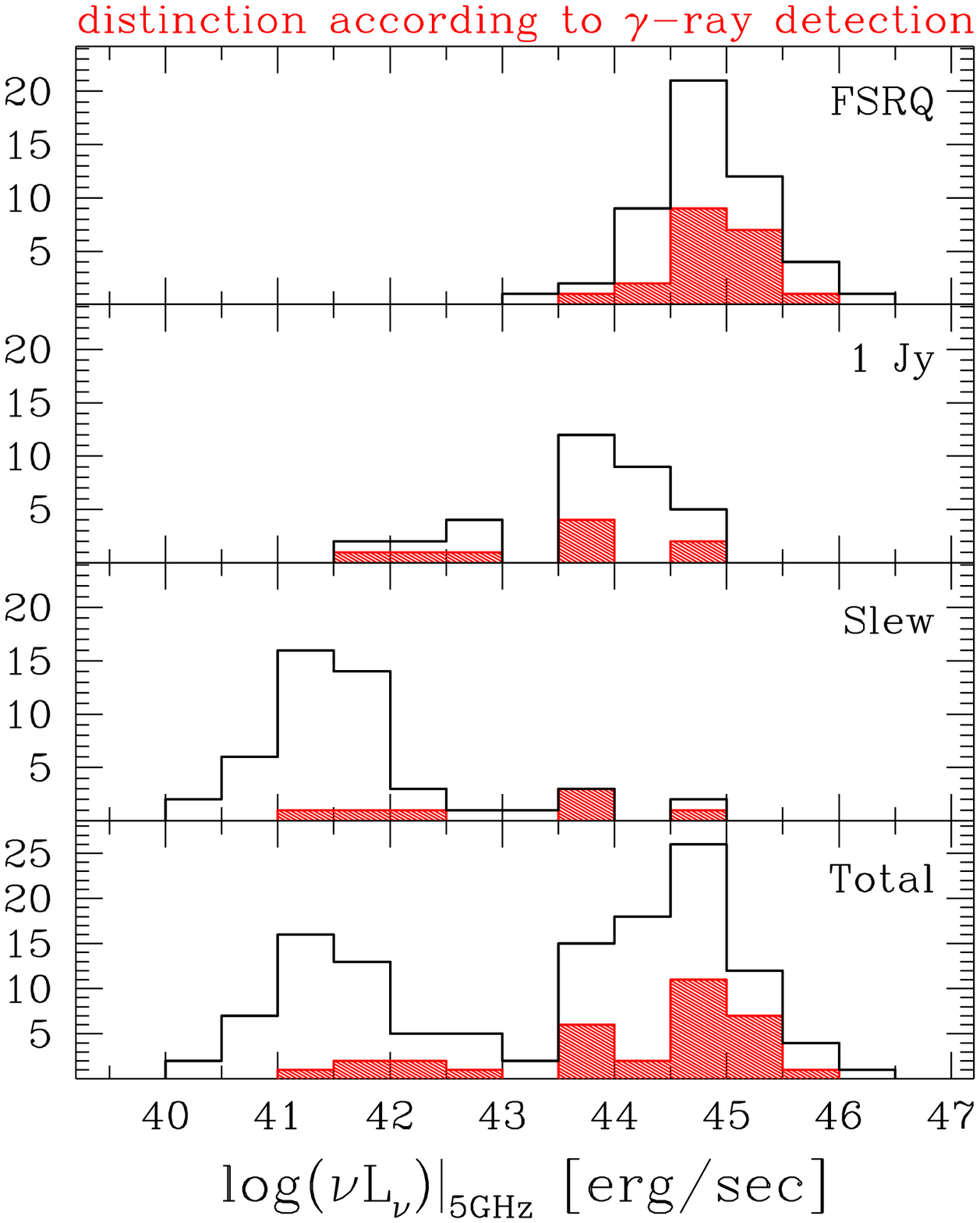,width=8.5truecm,height=8.5truecm}
\caption[h]{Distributions of L$_{\rm 5GHz}$ for the three complete
samples and the ``total blazars sample". 
Grey areas indicate $\gamma$--ray detected objects.}
\end{figure}

\begin{figure}
\psfig{file=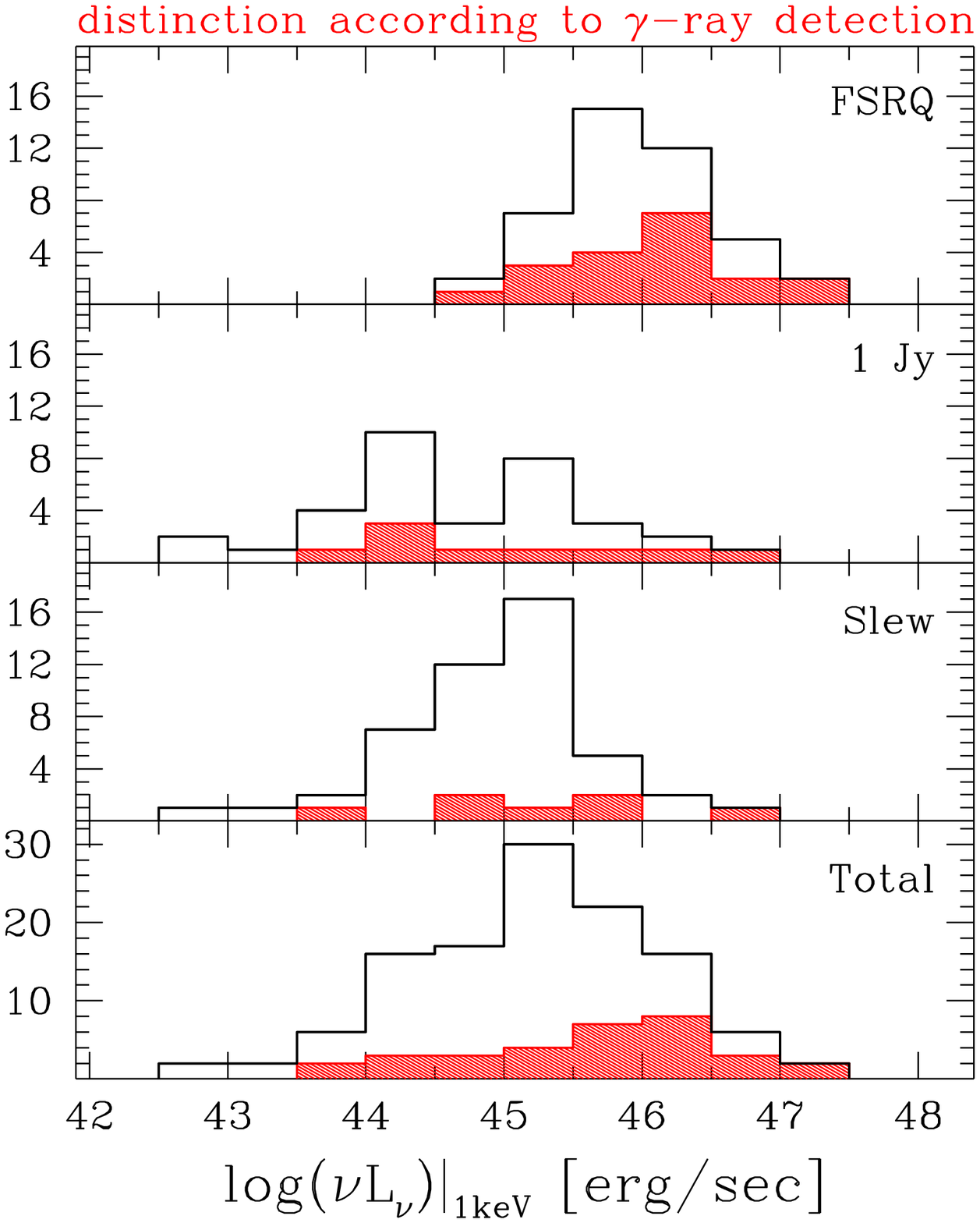,width=8.5truecm,height=8.5truecm}
\caption[h]{Distributions of L$_{\rm 1keV}$ for the three complete samples 
and the ``total blazars sample".
Grey areas indicate $\gamma$--ray detected objects.}
\end{figure}

\begin{figure}
\psfig{file=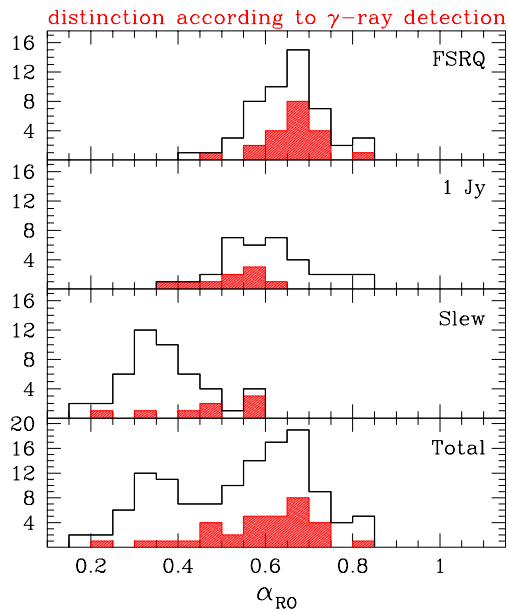,width=8.5truecm,height=8.5truecm}
\caption[h]{$\alpha_{\rm RO}$ distributions for the three complete samples 
and the ``total blazars sample".
Grey areas indicate $\gamma$--ray detected objects.}
\end{figure}

\begin{figure}
\psfig{file=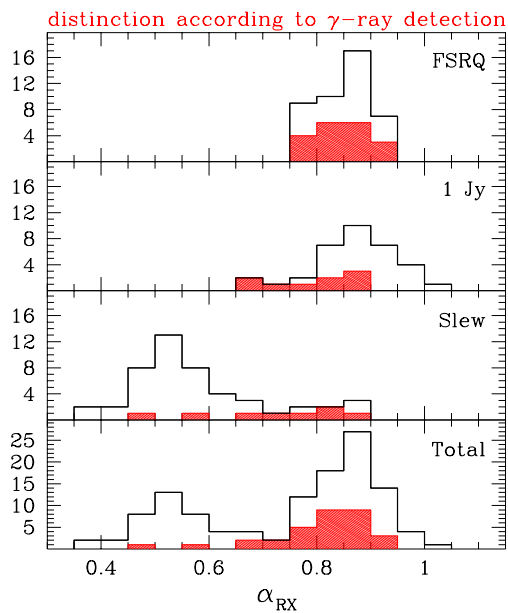,width=8.5truecm,height=8.5truecm}
\caption[h]{$\alpha_{\rm RX}$ distributions. As in Fig.~4.}
\end{figure}

The redshift distributions (Fig.~1) of the three complete samples show
the known tendency towards the detection of FSRQs at higher $z$,
being the latter ones more powerful radio sources, as shown in Fig.~2.  

In the same figure and in Fig.~3, the tendency for XBL to have similar
X--ray but lower radio luminosities compared to RBL is also apparent.
Correspondingly, it appears from Figs.~4,5  that $\alpha_{\rm RO}$ and 
$\alpha_{\rm RX}$ increase from XBL to RBL while FSRQ have 
$\alpha_{\rm RO}$ slightly larger and $\alpha_{\rm RX}$ similar to RBLs.  
Later on (section 3.2) we will show that there is a relationship 
between these spectral indices and the peak frequency of the
synchrotron component.

We note that for all of the distributions there is continuity in properties 
not only between the two BL Lac samples, but also between BL Lacs and FSRQs.

It is clear from Figs.~1-5 that for the RBL and FSRQ samples the 
$\gamma$--ray detected sources
do not differ from non--detected ones in any of the considered
quantities while for the Slew sample there is a tendency for
$\gamma$--ray loud sources to have larger L$_{\rm 5GHz}$,
$\alpha_{\rm RX}$ and $\alpha_{\rm RO}$. This indicates that either the radio 
luminosity or radio--loudness are important in determining the $\gamma$--ray 
detection.
On the contrary, and in some sense surprisingly, the X--ray luminosity
does not seem to play an important role with respect to the
$\gamma$--ray emission, although the X-ray band is the closest in energy
to the $\gamma$--rays .  
We will come back to this issue later on (\S 3.5). 
We checked the possible difference of means and variances of the distributions 
with the t--student's test and only for the $\alpha_{\rm RX}$ of the Slew 
sample the significance is higher than 95 per cent.

Except for the case of the Slew survey, we conclude that
the $\gamma$--ray detected sources are representative of the samples as
a whole, being indistinguishable from the others in terms of radio to
X-ray broad band properties and power.  

We also checked that the $\gamma$-ray detected sources belonging to our 
samples are homogeneous with respect to all of the $\gamma$-ray blazars 
detected so far.  In Fig.~6  we compare the redshift distributions, the radio 
luminosities and fluxes and the $\gamma$--ray spectral index. 
The grey shaded areas represent the $\gamma$--ray sources belonging to the
complete samples considered here. We conclude that there is no significant
difference.  

\begin{figure}
\psfig{file=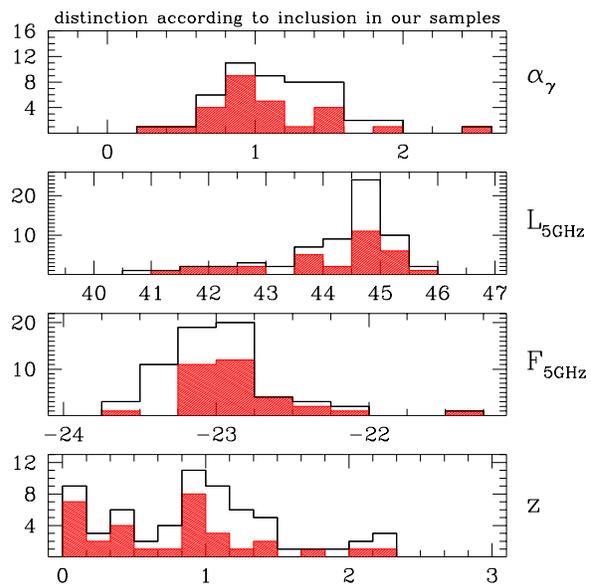,width=8.5truecm,height=8.5truecm}
\caption[h] {Distributions of redshift $z$, L$_{\rm 5GHz}$, radio flux
F$_{\rm 5GHz}$ and $\gamma$--ray spectral index
$\alpha_{\gamma}$ for the $\gamma$--ray loud sources detected by
EGRET, where the shaded ones refer only to sources in our three
complete blazar samples.}
\end{figure}

Nevertheless, we are aware that the limited sensitivity of the EGRET
instrument implies that at a given radio flux, 
only the $\gamma$--ray loudest sources are detected. Therefore the non detected
ones are probably on average weaker in $\gamma$-rays.
Impey (1996) quantified this effect by taking into account the correlation 
between radio and $\gamma$--ray luminosities 
(see \S 3.3.1), and  other observables. Assuming a Gaussian distribution of the
$\gamma$--ray to radio flux ratio he  estimated  the width of the distribution
and the "true" ratio referring to the whole population, which  could be a
factor 10 lower than the observed one.  
There could be a real spread in the intrinsic properties
of the blazar population,  the $\gamma$-ray detected blazars being
intrinsically louder than the rest of the population.
Alternatively this may be due to variability, i.e. a source is detected
only when it undergoes a flare. 
The observed ratio would thus refer to flaring states,  
while the average level of each source would be lower. 

Since variability is a distinctive property of blazars and has been observed 
to occur also in $\gamma$--rays, often with extremely large amplitude 
(greater than a factor 10) (\eg 3C 279, Maraschi et al 1994; PKS 0528+134, 
Mukherjee et al. 1996; 1622$-$297, Mattox et al. 1997) the latter
alternative is likely although the problem remains an open one.
We conclude that the average $\gamma$--ray luminosities computed here are 
necessarily overestimated.
However we chose not to correct for this effect given the uncertainties.
In particular the  "bias factor"  for different classes of blazars could 
be different if their $\gamma$--ray variability properties (amplitude and 
duty cycle) are (\eg Ulrich, Maraschi \& Urry 1997).

\subsection{Synchrotron peak frequency}

As previously noted the SEDs clearly show a broad peak between radio and 
UV--X--rays. 
In order to determine the position of the peak of the synchrotron
component in individual objects with an objective procedure,
we fitted the data points for each source
(in a $\nu$ vs. $\nu$L$_\nu$ diagram) with a third degree polynomial,
which yields a  complex SED profile, with an upturn allowing for X--ray 
data points not to lay on the direct extrapolation from the lower 
energy spectrum.  
In many cases there is evidence that the X--ray component, even in the
soft {\it ROSAT} PSPC band, is due to the inverse Compton process (e.g.
Sambruna 1997; Comastri et al. 1997). Thus to impose that the X--ray point
smoothly connects to the lower energy data, as would happen in a parabolic
fit, could be misleading for a determination of the synchrotron peak
frequency.  
We used a simple parabola when the cubic fit was not able to find a maximum, 
which typically happens when the peak occurs at energies higher than X--rays.  
In fact when the peak moves to high enough frequencies
(typically beyond the IR band), the  X--ray flux is completely
dominated by the synchrotron emission, and  the results given by
the cubic and parabolic fits are fully consistent.
In 8 cases neither the cubic nor the parabolic fit were able to determine a 
peak frequency/luminosity mainly due to the paucity of data points.

\begin{figure}
{\vbox{
\psfig{file=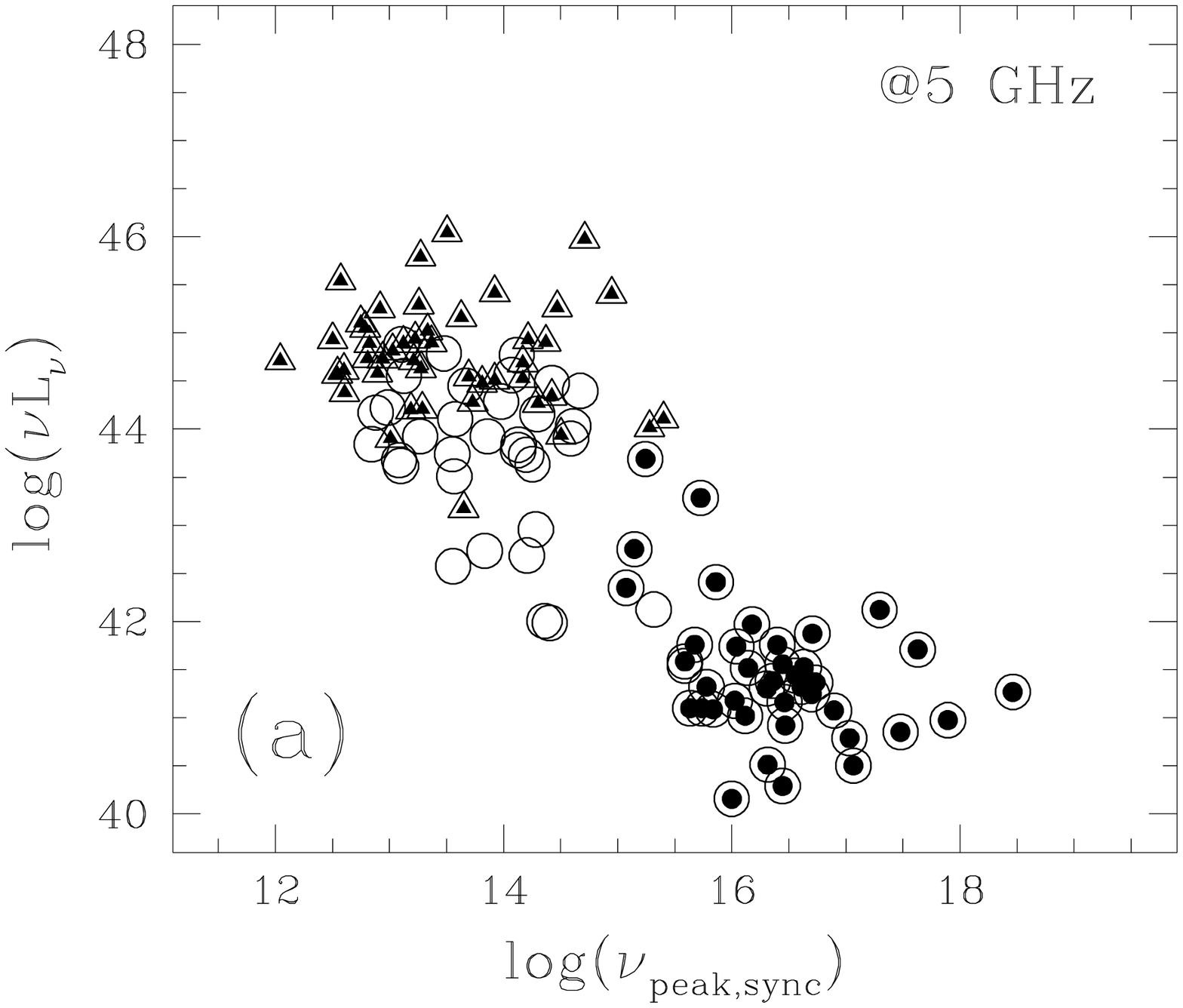,width=8.5truecm,height=8.5truecm,rheight=7.3truecm}
\psfig{file=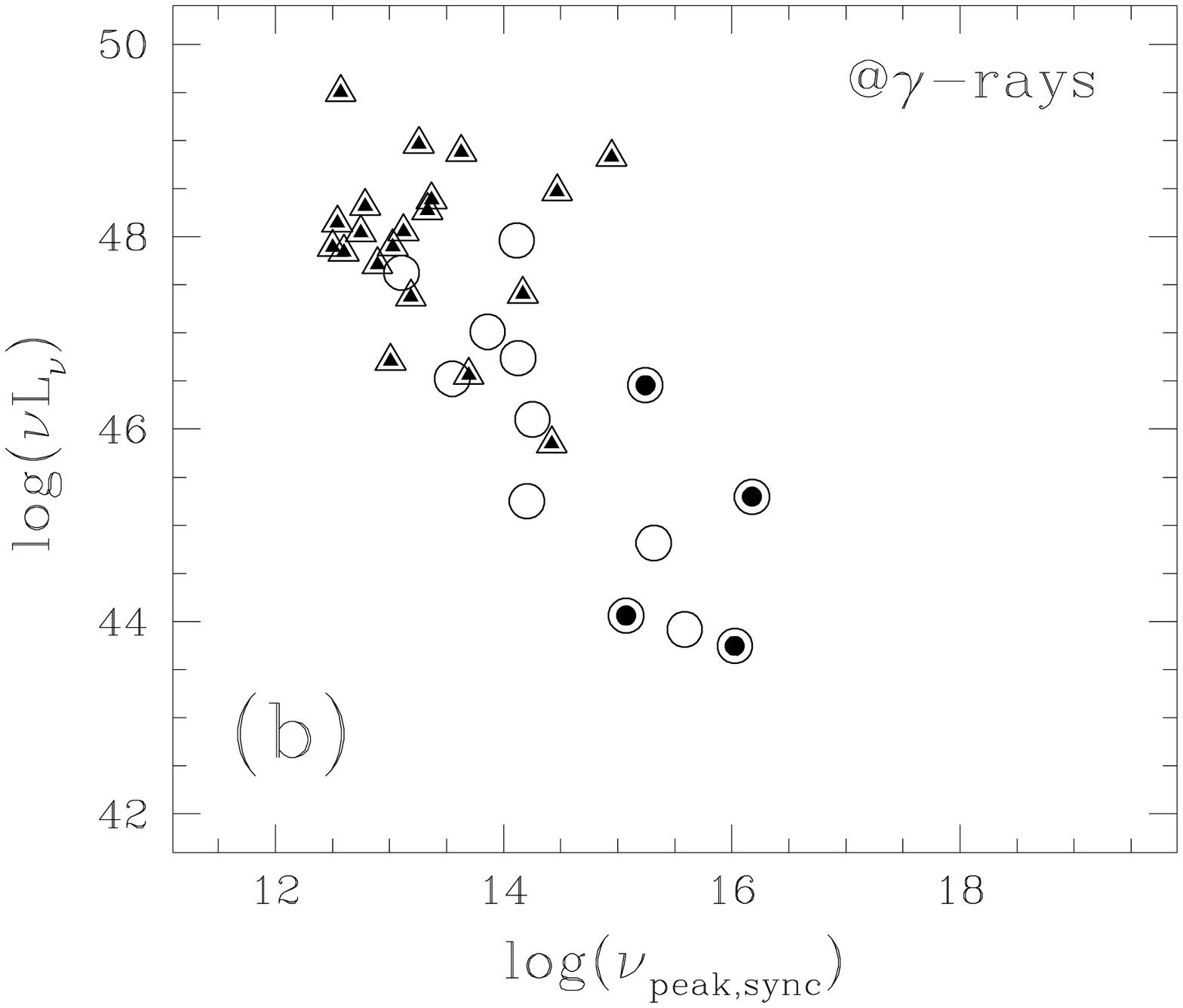,width=8.5truecm,height=8.5truecm,rheight=7.3truecm}
\psfig{file=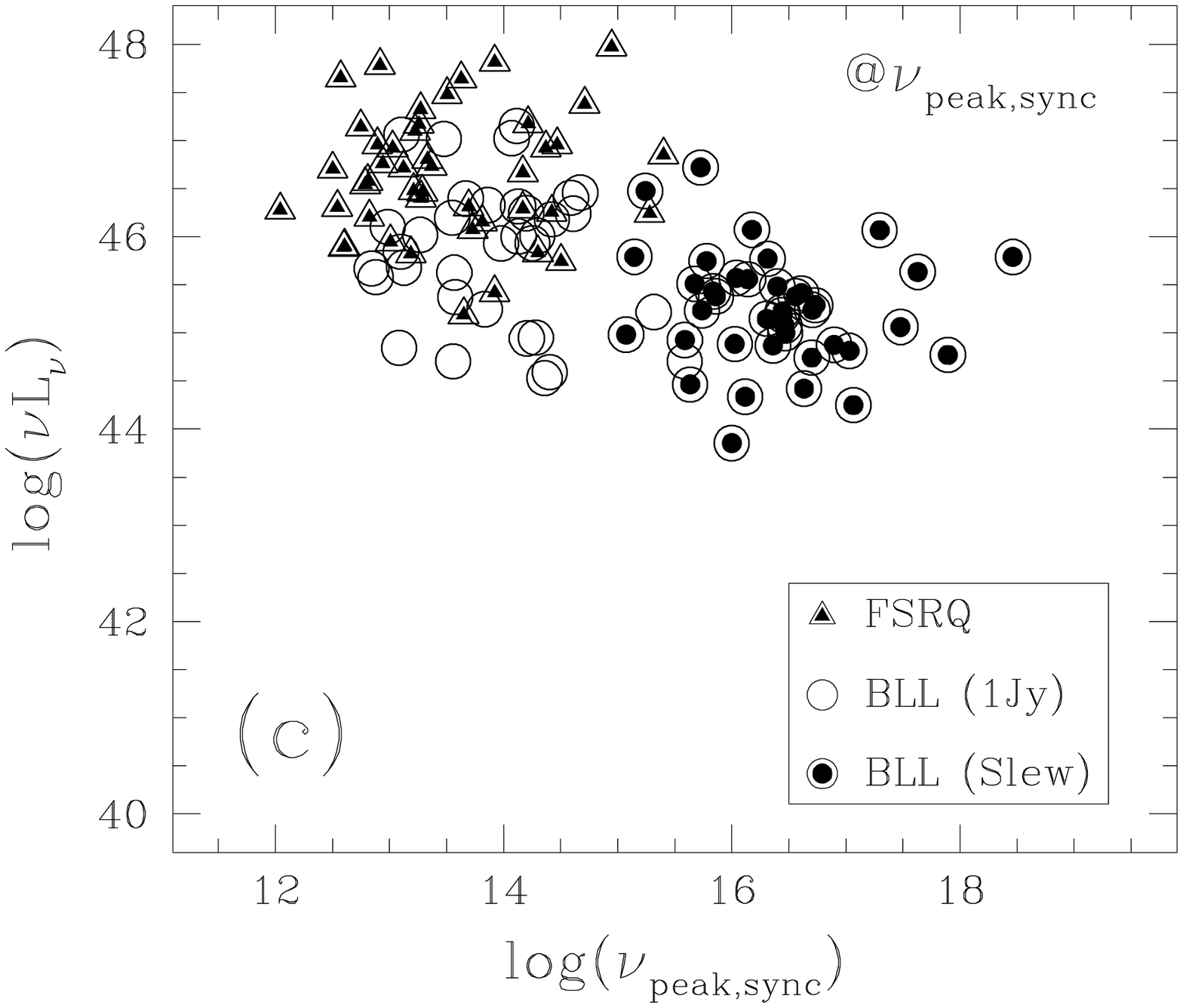,width=8.5truecm,height=8.5truecm,rheight=7.3truecm}
}}
\caption[h] {The peak frequency of the synchrotron component, 
$\nu_{\rm peak,sync}$, as derived with the polynomial fits, plotted against 
a) the radio luminosity L$_{\rm 5GHz}$,
b) the $\gamma$--ray luminosity L$_\gamma$, and
c) the fitted peak luminosity of the synchrotron 
component, L$_{\rm peak,sync}$.}
\end{figure}

\subsubsection{Synchrotron Peak Frequency vs. Luminosity}

The peak frequencies derived with the above procedure (defined as
the frequencies of the maximum in the fitted polynomial function) are
plotted in Fig.~7a,b,c versus the radio and $\gamma$--ray luminosities
and  vs the corresponding peak luminosities, as determined  from the
fits. Let us stress once again the continuity between the different samples.
Considering the samples together strong correlations are present between 
these quantities, in the sense of $\nu_{\rm peak,sync}$ decreasing with 
increasing luminosity. 
The results of Kendall's $\tau$ statistical test (Table 4) show that the
correlations are highly significant. 

Since on one hand in flux limited samples spurious correlations can be
introduced by the luminosity/redshift relation and on the other hand the
correlations might be due to evolutionary effects genuinely related to
redshift, we checked its role in two ways. We estimated the
possible correlation of the relevant quantities with redshift directly, and 
performed  partial correlation tests between two quantities subtracting out 
the common dependence on $z$ (Padovani 1992b)(see results in Table~4).
In addition, in order to have an independent check on the redshift bias,
we also considered the significance of the correlations restricting them
to objects with $z < 0.5$ (see Table 4).  

The correlation between $\nu_{\rm peak,sync}$ and L$_{\rm 5 GHz}$
still holds after subtraction of the very strong dependence on redshift. 
The same is true for the relation between $\nu_{\rm peak,sync}$ and 
the $\gamma$-ray luminosity, although the significance is much smaller,
due to the smaller number of sources.
On the other hand the correlation between $\nu_{\rm peak,sync}$ and 
L$_{\rm peak,sync}$ is strongly weakened when subtracting the redshift effect.

Considering only the $z < 0.5$ interval the significance of the first two 
correlations persists and does not change when the redshift 
dependence is subtracted.
These values can then be considered as irreducible, being the
signature of a {\it true dependence} of $\nu_{\rm peak,sync}$ on luminosity.
This result can also be read as a check of the reliability of the partial 
correlation procedure.
On the contrary  the correlation $\nu_{\rm peak,sync}$ vs. L$_{\rm peak,sync}$ 
disappears at low redshifts, due to the narrow interval of values
spanned by L$_{\rm peak,sync}$, that varies less with the change
of peak frequency than both radio and $\gamma$--ray luminosity do.

\begin{table*}
\caption{Correlation probabilities according to the Kendall's $\tau$ test.  
(1--2): quantities considered in the correlation;  
(3--10): significance of the ``null--hypothesis'', \ie that the 
correlation is the result of pure chance, for the various quantities: 
in columns (3--6) for tests performed without any redshift constraint,
while in (7--10) results for tests taking into account only sources with
z $\gs$ 0.5.
In (3),(7) the ``face value" correlation between x$_1$ and x$_2$.
In (4--5) and (8--9) the significance of the correlation of each
quantity x$_1$ and x$_2$ with redshift.
In (6) and (10) the ``net" x$_1$/x$_2$ correlation remaining from
(3) and (7) after subtraction of the redshift dependence of x$_1$ and x$_2$ 
via partial correlation algorithm.  A dash is reported the correlation is
not significant (\ie probability $<$ 90 \%.} 
\begin{tabular}{@{}ccccccccccccccc} 
\hline
& & && 
\multicolumn{5}{c}{z unconstrained (\# 109$^{a}$)}  && 
\multicolumn{5}{c}{z $<$ 0.5 (\# 51$^{a}$)} \\
\cline{5-9} \cline{11-15} \\
&{x$_1$} & {x$_2$} && 
{x$_1$/x$_2$} && {x$_1$/z} & {x$_2$/z} & {x$_1$/x$_2$$-Z$} && 
{x$_1$/x$_2$} && {x$_1$/z} & {x$_2$/z} & {x$_1$/x$_2$$-Z$} \\ 
&(1)&(2)&&
(3)&&(4)&(5)&(6)&&
(7)&&(8)&(9)&(10) \\
\hline
&&&&&&&&&&&&&&\\
&$\nu_{\rm peak, sync}$ & L$_{\rm 5 GHz}$ && 
1.2e-16 && [1.1e-9] & [1.9e-31] & 1.1e-11 && 
1.3e-8  && [ --- ]  & [4.0e-4]  & 9.2e-9  \\
&&&&&&&&&&&&&&\\
&$\nu_{\rm peak, sync}$ & L$_{\rm peak, sync}$ && 
4.9e-8 && [1.1e-9] & [9.5e-27] & 6.5e-2 && 
 ---   && [ --- ]  & [1.2e-7]  & ----  \\
&&&&&&&&&&&&&&\\
&$\nu_{\rm peak, sync}$ & $\alpha_{\rm RO}$ &&  
1.1e-23 && [1.1e-9] & [7.7e-11] & 4.7e-19 && 
1.8e-7  && [ --- ]  & [ --- ]   & 2.9e-7 \\
&&&&&&&&&&&&&&\\
&$\nu_{\rm peak, sync}$ & $\alpha_{\rm RX}$ &&  
4.2e-17 && [1.1e-9] & [6.0e-5] & 1.9e-14 && 
4.1e-15 && [ --- ]  & [ --- ]  & 8.4e-15 \\
&&&&&&&&&&&&&&\\
\hline
\multicolumn{15}{c}{Correlations involving $\gamma$--ray data} \\
\hline
\multicolumn{5}{l}{\underbar{EGRET sources included in our samples}} &&&&&&&&&\\
& & && 
\multicolumn{5}{c}{z unconstrained (\# 31)}  && 
\multicolumn{5}{c}{z $<$ 0.5 (\# 11)} \\
\cline{5-9} \cline{11-15} \\
&$\nu_{\rm peak, sync}$ & L$_\gamma$ &&  
2.1e-3 && [3.4e-2] & [1.5e-10] & 1.8e-2 && 
2.4e-2 && [ --- ]  & [2.4e-3]  & 6.4e-2 \\
&&&&&&&&&&&&&&\\
&L$_{\rm 5GHz}$ & L$_\gamma$ &&  
6.0e-11 && [5.5e-10] & [1.5e-10] & 4.2e-5 && 
3.9e-3  && [5.2e-2]  & [2.4e-3]  & 2.4e-2 \\
&&&&&&&&&&&&&&\\
&$\nu_{\rm peak, sync}$ & L$_\gamma$/L$_{\rm peak,sync}$ &&  
2.7e-5 && [3.4e-2] & [1.5e-5] & 2.2e-4 && 
1.4e-3 && [ --- ]  & [7.3e-1] & 4.1e-3  \\
&&&&&&&&&&&&&&\\
&$\nu_{\rm peak, sync}$ & L$_\gamma$/L$_{\rm 5500\AA}$ &&  
2.5e-6 && [3.4e-2] & [4.2e-5] & 1.7e-5 && 
2.4e-3 && [ --- ]  & [ --- ]  & 6.8e-3 \\
&&&&&&&&&&&&&&\\
\hline
\multicolumn{5}{l}{\underbar{sources in the whole EGRET sample}} &&&&&&&&& \\
& & && 
\multicolumn{5}{c}{z unconstrained (\# 60)}  && 
\multicolumn{5}{c}{z $<$ 0.5 (\# 15)} \\
\cline{5-9} \cline{11-15} \\
&L$_{\rm 5GHz}$ & L$_\gamma$ &&  
4.8e-15 && [4.8e-15] & [1.8e-14] & 2.4e-6 && 
1.8e-3  && [1.1e-2]  & [1.4e-4]  & 4.1e-2 \\
&&&&&&&&&&&&&&\\
\hline
\end{tabular}
{\raggedright 
{\bf Notes to Table 4:}\par\small
\noindent$^{(a)}$ we considered only sources with at least a lower limit 
on redshift, and for which it has been possible to determine the 
``synchrotron" peak frequency by means of the polynomial fit. \par
}
\medskip
\end{table*}

\begin{figure}
\psfig{file=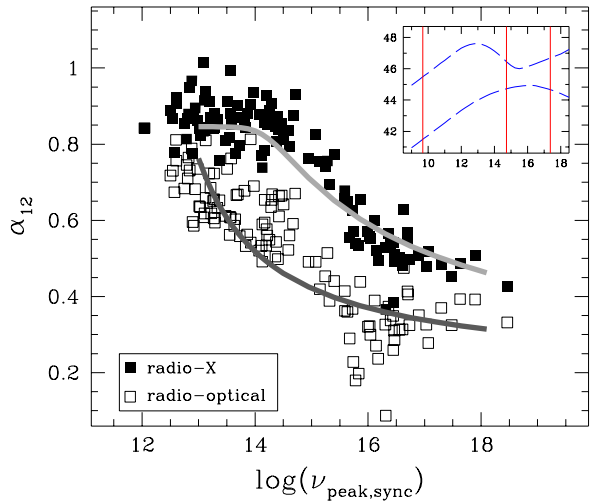,width=8.5truecm,height=8.5truecm,rheight=7.8truecm}
\caption[h] {The broad band spectral indices $\alpha_{\rm RO}$ and
$\alpha_{\rm RX}$ are plotted vs $\nu_{\rm peak,sync}$.
The curved lines overlayed to the data points represent the relations
$\alpha_{\rm RO}$--$\nu_{\rm peak,sync}$ and 
$\alpha_{\rm RX}$--$\nu_{\rm peak,sync}$ obtained from a ``synthetic" set 
of SEDs. 
Details on the adopted analytical parameterization are given in text, \S 4.
Two examples of typical SEDs are reported for reference in the inset: 
the top one peaking at $\nu \sim 10^{13}$ Hz, with
$\alpha_{\rm RO} = 0.76$ and $\alpha_{\rm RX} = 0.85$, the bottom one at
$\nu \sim 10^{16}$ Hz ($\alpha_{\rm RO} = 0.35$ and $\alpha_{\rm RX} = 0.56$).
The three vertical lines mark the frequencies corresponding to 5 GHz, 
5500 \AA, and 1 keV, entering in the definitions of $\alpha_{\rm RO}$ and
$\alpha_{\rm RX}$.
}
\end{figure}

\subsubsection{Synchrotron Peak Frequency vs. Broad Band Spectral indices}

 The relations between the synchrotron peak frequency and each of
the two point spectral indices $\alpha_{\rm RO}$ and $\alpha_{\rm RX}$
are shown in Fig.~8.  
Also these quantities are strongly correlated (see Table~4). In fact  
recent papers (e.g. Maraschi et al. 1995; Comastri et al. 1995; 
Comastri et al. 1997), suggested that the position of the synchrotron peak 
could be devised from the values of broad band spectral indices.

We see that the knowledge of any of the two
spectral indices is enough to guess the position of the peak of the
synchrotron component, except for some ranges, namely $\nu_{\rm
peak,sync} > 10^{16-17}$ Hz for both $\alpha_{\rm RO}$ and
$\alpha_{\rm RX}$, and $\nu_{\rm peak,sync} < 10^{14}$ Hz for
$\alpha_{\rm RX}$.  These ``failures'' can be explained bearing in
mind the typical shape of the blazar SEDs (see inset in Fig.~8): 
when the spectrum peaks at
low frequencies, X--rays are typically dominated by the inverse
Compton, flat spectrum, component whose luminosity level is strongly
correlated with the radio one (Fossati et al. 1997), and then the
X--ray/radio ratio (i.e. $\alpha_{\rm RX}$) tends to a fixed value.
Conversely the Compton component begins to dominate the ({\it ROSAT})
 X-ray band when $\alpha_{\rm RX}
\sim 0.75$, corresponding to $\nu_{\rm peak,sync} \lg 3 \times 10^{14}$ Hz.  
It is interesting to note that the adopted dividing threshold between LBL and
HBL has been set to this same value from purely practical purposes, while 
 in the light of the result above it assumes a more ``physical" meaning. 
LBL sources have Compton--dominated soft--X--ray emission, while in HBL 
this is pure synchrotron.
 
At the other end of the spectrum a problem arises when
 $\nu_{\rm peak,sync}$ moves at energies higher than that used
to compute the broad band spectral index.  The reason is that the
ratio between, for instance, optical and radio luminosity is no
longer sensitive to the peak moving further towards higher frequencies,
because both the radio and optical bands lay on the same (rising)
branch of the synchrotron ``bump''.

For comparison we draw in Fig.~8  the loci of 
$\alpha_{\rm RO}$--$\nu_{\rm peak,sync}$ and 
$\alpha_{\rm RX}$--$\nu_{\rm peak,sync}$ obtained from a set of SEDs 
of the kind reported in the inset, and that we are going to discuss in
more detail in section 3.5. 
The parameterization describes the observed features very well.

\begin{figure}
\psfig{file=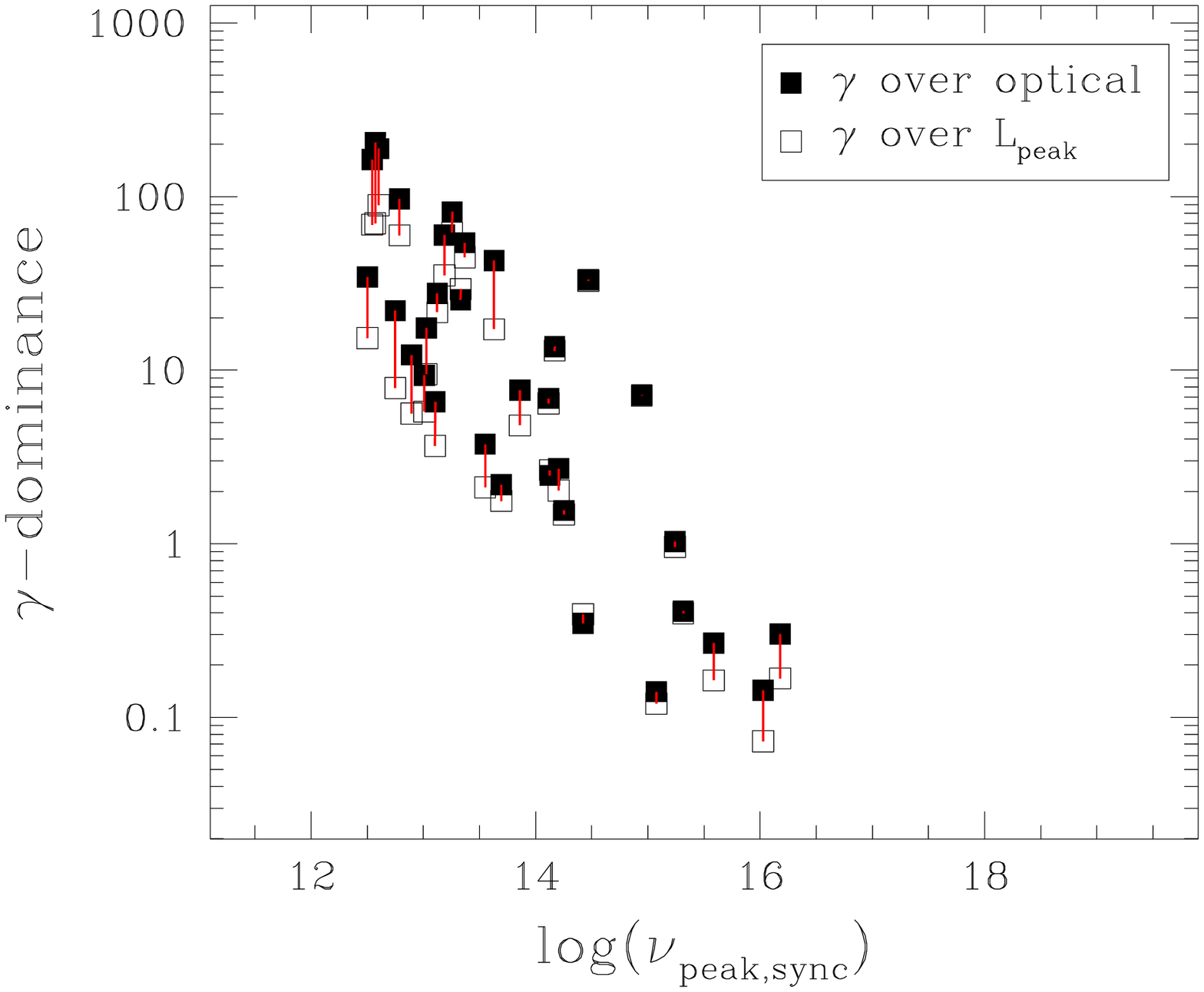,width=8.5truecm,height=8.5truecm,rheight=7.5truecm}
\caption[h] {The $\gamma$--ray dominance (according to two
definitions, see text) versus the synchrotron peak frequency 
$\nu_{\rm peak,sync}$}
\end{figure}

\subsubsection{Synchrotron Peak Frequency vs. $\gamma$--ray dominance}

In Fig.~9 $\nu_{\rm peak,sync}$ is plotted against the $\gamma$--ray
dominance parameter, defined as the ratio between the $\gamma$--ray and
the synchrotron peak luminosities. 
A strong correlation (see Table~4) is present over four orders of magnitude 
in $\nu_{\rm peak,sync}$, in the sense of a decrease in the $\gamma$--ray dominance
with an increase of the synchrotron peak frequency.  
 In the same figure we plotted also the ratio  between the $\gamma$--ray and optical 
luminosities, to check if the latter
could eventually be a good indicator of the $\gamma$--ray
dominance, with the advantage of being an observed quantity. In fact there is  
little difference, at most a factor 3 for a quantity spanning more
than three decades.

\subsection{Average SEDS}

Having discussed extensively the possible biases introduced by the limited
number of $\gamma$--ray detected sources in the complete samples we
construct here the  average SEDs for each sample. We will come back later
to the bias problem (Section 4).

The averaging procedure has been performed on the logarithms of
the luminosities at each frequency.  
Apart from the problems in the $\gamma$-ray range discussed above the
incompleteness of the data coverage at some frequencies could also
introduce a bias in the average values.  
For instance in the Slew survey sample only 10/48 objects have a flux 
measured at 230 GHz, and they are the more luminous sources at 5 GHz.  
Averaging independently L$_{\rm 230GHz}$ (for 10 objects) and L$_{\rm 5GHz}$ 
(for 48 objects) we would obtain a ratio between the two luminosities higher 
than that derived considering only the subsample of 10 sources, 
and  presumably higher than the actual one, too.

To reduce this kind of effect we first normalized the monochromatic 
luminosities to the radio luminosity for each source, we 
computed average ratios
$\langle \log(L_{\nu^*}/L_{\rm 5GHz}) \rangle|_{sub}$, considering only the 
subsample of sources with a measured flux at $\nu^*$, and used that ratio
to compute the average monochromatic luminosity at $\nu*$ for all sources
in the sample as $\langle \log(L_{\nu^*}) \rangle|_{all} = 
\langle \log(L_{\rm 5GHz}) \rangle|_{all} + 
\langle \log(L_{\nu^*}/L_{\rm 5GHz}) \rangle|_{sub}$.  
In this way we basically averaged  the spectral
shape between $\nu^*$ and 5 GHz for the measured objects and assigned that
spectral shape to the sample. 

The X--ray and $\gamma$--ray spectral indices have been averaged with
a simple mean, without weighting.

\begin{figure}
\psfig{file=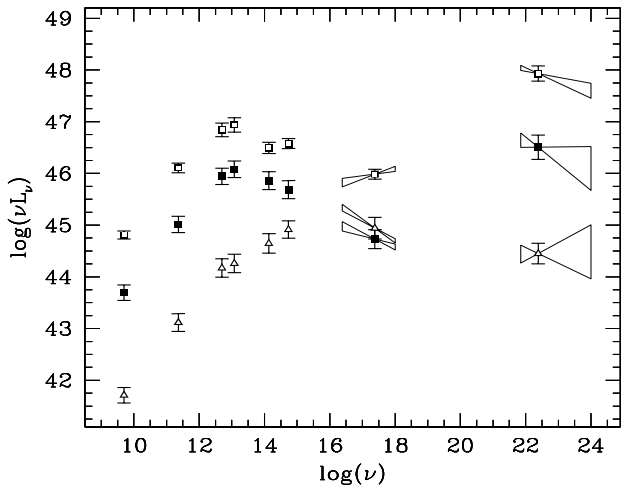,width=8.5truecm,height=8.5truecm,rheight=7.8truecm}
\caption[h] {The average SEDs for each of the samples are shown.
From top to bottom (referring to radio luminosity) Wall \& Peacock
FSRQs (empty boxes), 1 Jy BL Lac sample (filled boxes) and Slew survey BL 
Lac sample (triangles) . 
These latter two are in reversed order in X--ray band, the lowest spectrum 
being that of the 1 Jy sample.}
\end{figure}

\begin{figure}
\psfig{file=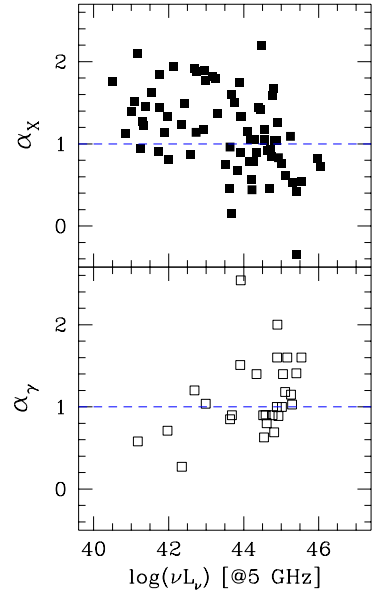,width=8.5truecm,height=8.5truecm}
\caption[h] {X--ray and $\gamma$--ray spectral indices plotted against 
radio luminosity.}
\end{figure}

The average broad band spectra for each of the three samples are shown in
Fig.~10.  The 6 sources common to the radio and the X--ray selected BL Lac
samples are considered in both of them.  Average luminosities entering
Fig.~10 are reported in Table~5 together with the number of sources
contributing at each frequency. 

It is apparent from Fig.~10 that the three samples refer to objects with
different average integrated luminosities and that the peak frequency
of the power emitted between the radio and the X-ray band moves from
the X-ray to the far infrared band going from the XBL to the FSRQ
samples as anticipated from the analysis of single objects in the
previous section 3.2. Correspondingly the $\gamma$--ray luminosities
increase and the $\gamma$--ray spectra steepen suggesting that also the
peak frequency of the high energy emission moves to lower frequencies.
The overall similarity and regularity of the SEDs of the different samples
as well as the continuity in the properties of the individual objects
discussed in section 3.2 suggest a basic similarity of all blazars
irrespective of their original classification and different appearance
in a specific spectral band.

\begin{table*}
\caption{Average luminosities, for each sample and for the total one 
(divided in bins of radio luminosity)}
\begin{tabular}{@{}clccccccccc } 
\hline
&&\multicolumn{3}{c}{Complete Samples}&& \multicolumn{5}{c}{Total Blazar Sample (log(L$_{\rm 5GHz}$) intervals)} \\
\cline{3-5}\cline{7-11} \\     
& Band & Slew & 1 Jy & W\&P && $<$42 & 42$-$43 & 43$-$44 & 44$-$45 & $>$45 \\
\hline
	&&&&&&&&&&\\
& 5 GHz     & 41.71 & 43.69 & 44.81 && 41.24 & 42.47 & 43.71 & 44.54 & 45.39 \\
&           &  48   &  34   &  50   &&  38   &  10   &  17   &  44   &  17   \\
	&&&&&&&&&&\\
& 230 GHz   & 43.11 & 45.01 & 46.11 && 42.64 & 43.77 & 45.13 & 45.83 & 46.63 \\
&           &  10   &  34   &  50   &&   5   &   7   &  15   &  44   &  17   \\
	&&&&&&&&&&\\
& 60 $\mu$m & 44.17 & 45.94 & 46.84 && 43.73 & 44.65 & 46.09 & 46.65 & 47.61 \\
&           &  12   &  19   &  13   &&   6   &   5   &   8   &  16   &   2   \\
	&&&&&&&&&&\\
& 25 $\mu$m & 44.25 & 46.07 & 46.93 && 43.74 & 44.95 & 46.08 & 46.79 & 47.69 \\
&           &  10   &  15   &   8   &&   4   &   6   &   7   &   9   &   2   \\
	&&&&&&&&&&\\
& K--band   & 44.64 & 45.86 & 46.49 && 44.42 & 45.04 & 45.96 & 46.27 & 47.21 \\
&           &  23   &  31   &  28   &&  13   &  10   &  15   &  32   &   6   \\
	&&&&&&&&&&\\
& V--band   & 44.91 & 45.68 & 46.58 && 44.61 & 45.01 & 45.82 & 46.27 & 47.21 \\
&           &  48   &  34   &  50   &&  38   &  10   &  17   &  44   &  17   \\
	&&&&&&&&&&\\
& 1 keV     & 44.94 & 44.72 & 45.98 && 44.81 & 44.11 & 44.92 & 45.66 & 46.50 \\
&           &  48   &  32   &  43   &&  38   &  10   &  15   &  42   &  12   \\
	&&&&&&&&&&\\
& 100 MeV   & 44.45 & 46.50 & 47.93 && 44.24 & 44.79 & 46.67 & 47.71 & 48.68 \\
&           &   7   &   8   &  20   &&   3   &   5   &   9   &  33   &  12   \\
	&&&&&&&&&&\\
& $\alpha_{\rm X}$  & 1.40 $\pm$ 0.07 & 1.25 $\pm$ 0.09 & 0.83 $\pm$ 0.08 && 1.37 $\pm$ 0.09 & 1.55 $\pm$ 0.15 & 1.16$\pm$ 0.14& 1.11 $\pm$ 0.08 & 0.57 $\pm$ 0.13 \\
&           &  24   &  31   &  24   &&  16   &   8   &  14   &  26   &   9   \\
	&&&&&&&&&&\\
& $\alpha_\gamma$& 0.98 $\pm$ 0.32 & 1.26 $\pm$ 0.26 & 1.21 $\pm$ 0.09 && 0.64 $\pm$ 0.07 & 0.73 $\pm$ 0.47 & 1.37 $\pm$ 0.31 & 1.06 $\pm$ 0.13 & 1.30 $\pm$ 0.08 \\
&           &   6   &   6   &  18   &&   2   &   4   &   7   &  25   &  11   \\
	&&&&&&&&&&\\
\hline
\end{tabular}
{\raggedright
{\bf References to Table 5:}\par\small
(5 and 230 GHz): 
Becker, White \& Edwards 1991;       
Bloom et al. 1994;                   
Gear et al. 1986;                    
Gear 1993a;                          
Gear et al. 1994;                    
K\"uhr et al. 1981;                  
K\"uhr \& Schmidt 1990;              
Perlman et al. 1996a;                
Reuter et al. 1997;                  
Steppe et al. 1988, 1992, 1993;      
Stevens et al. 1994;                 
Stickel et al. 1991;                 
Stickel et al. 1993;                 
Stickel, Meisenheimer \& K\"uhr 1994;
Tornikovski et al. 1993, 1996;       
Terasranta et al. 1992.              
\par
(IR--optical data):
Allen, Ward \& Hyland 1982;          
Ballard et al. 1990;                 
Bersanelli et al. 1992;              
Bloom et al. 1994;                   
Brindle et al. 1986;                 
Brown et al. 1989;                   
Elvis et al. 1994;                   
Falomo et al. 1988;                  
Falomo et al. 1993a;                 
Falomo et al. 1993b;                 
Falomo, Scarpa \& Bersanelli  1994;  
Gear et al. 1986;                    
Gear 1993b;                          
Glass 1979, 1981;                    
Holmes et al. 1984;                  
Impey \& Brand 1981;                 
Impey \& Brand 1982;                 
Impey et al. 1982;                   
Impey et al. 1984;                   
Impey \& Neugebauer 1988;            
Impey \& Tapia 1988, 1990;           
Jannuzi, Smith \& Elston 1993, 1994; 
Landau et al. 1986;                  
Lepine, Braz \& Epchtein 1985;       
Lichtfield et al. 1994;              
Lorenzetti et al. 1990;              
Mead et al. 1990;                    
O'Dell et al. 1978;                  
Pian et al. 1994;                    
Sitko \& Sitko 1991;                 
Smith et al. 1987;                   
Stevens et al. 1994;                 
Wright, Ables \& Allen 1983.         
\par
(X--rays): 
Brinkmann et al. 1994;               
Brinkmann et al. 1995;               
Brunner et al. 1994;                 
Comastri et al 1995;                 
Comastri et al 1997;                 
Lamer, Brunner \& Staubert 1996;     
Maraschi et al. 1995;                
Perlman et al. 1996a;                
Perlman et al. 1996b;                
Sambruna 1997;                       
Urry et al. 1996.                    
\par
($\gamma$--rays):   
Bertsch et al. 1993;                 
Catanese et al. 1997;                
Chiang et al. 1995;                  
Dingus et al. 1996;                  
Fichtel et al. 1994;                 
Hartman et al. 1993;                 
Lin et al. 1995;                     
Lin et al. 1996;
Madejski et al. 1996;                
Mattox et al. 1997;                  
Mukherjee et al. 1995, 1996;         
Nolan et al. 1996;                   
Quinn et al. 1996;                   
Radecke et al. 1995;                 
Shrader et al. 1996;                 
Sreekumar et al. 1996;               
Thompson et al. 1993;                
Thompson et al. 1995;                
Thompson et al. 1996;                
Vestrand, Stacy \& Sreekumar 1995;   
von Montigny et al. 1995.            
\par
}
\medskip
\end{table*}


We therefore considered the merged total sample with the scope
of finding the key parameter(s) governing the whole blazar
phenomenology.
Since luminosity appears to have an important role in that it correlates
with the main  spectral parameters we decided to bin the total blazar sample
according to luminosity, irrespective of the original classification.
We used the 5 GHz radio luminosity which is available for all objects.
It may be desirable to use the total integrated luminosity which in
all cases is close to the $\gamma$--ray one. However the latter is only 
available for a few objects.
We note that a correlation between $\gamma$--ray and radio luminosity has been
claimed by many authors using different techniques 
(Dondi \& Ghisellini, 1995; Mattox et al. 1997). 
It is however still being debated whether it is true or it arises from
selection effects, connected with the common redshift dependence of 
luminosities, and with the exclusion of upper limits, which could favour
the appearance of a spurious correlation.
It is worth mentioning that M\"ucke et al. (1997) using a technique
designed to take into account both these effects did not find any
significant correlation between radio and $\gamma$-ray data for a sample
of 38 extragalactic EGRET sources.

We also checked this correlation on both the 31 EGRET detected sources
included in our samples and the larger ``comparison sample" of 62 EGRET sources.
In Table~4 we report the significance of the correlation, together with
its value after subtracting the common redshift dependence through a partial
correlation test, and its significance for samples restricted to $z<0.5$.
In all cases the radio and $\gamma$--ray luminosities correlate significantly.

In Figs.~11 $\alpha_{\rm X}$ and $\alpha_\gamma$ for individual sources are 
plotted against the radio power, both showing a good correlation with it. 
Comastri et al. (1997) discussed the interesting consequences of the apparent 
anti--correlation between X--ray and $\gamma$-ray spectral indices, without 
relating it to any ``absolute'' parameter, such as luminosity.
Here again we see that these other spectral properties have a dependence
on radio luminosity. 

\begin{figure}
\psfig{file=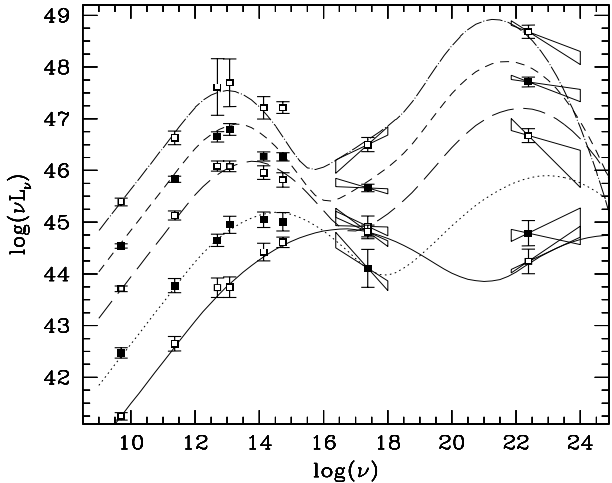,height=8.5truecm,width=9.0truecm,rheight=7.8truecm}
\caption[h] {Average SEDs for the ``total blazar sample'' binned  
according to radio luminosity irrespective of the original classification.
The overlayed curves are analytic approximations obtained according to
the one--parameter--family definition described in the text.}
\end{figure}

Since in some luminosity bins the number of $\gamma$-ray detected
sources is small, we used an indirect procedure to associate
$\gamma$--ray fluxes and spectra to our average SEDs, taking advantage
of the whole body of information regarding the $\gamma$--ray
properties of blazars.  Namely for each luminosity bin
$\langle$L$_{\gamma}\rangle$ and $\langle\alpha_\gamma\rangle$ were
computed from blazars in the general EGRET--detected sample falling
into the same L$_{\rm 5GHz}$ bin.  The basic assumption is the
uniformity of the spectral properties, as discussed in section 3.1.
The resulting SEDs are shown in Fig.~12 and average luminosities,
X--ray and $\gamma$--ray spectral indices, and number of sources are
reported in Table~5. The most interesting result is that the trends
pointed out for the three separate sub--classes of blazars (Fig.~10)
hold for the total blazar sample, irrespective of the original
classification of sources, when the radio luminosity is adopted as the
key parameter characterizing each object.

\section{Discussion}

In Fig.~12 we superimposed to the averaged data a set of (dashed)
lines, whose main goal is to guide the eye.  
The radio--X--ray SED is approximated with a power law starting in the
radio domain continuously connecting at $\nu \simeq 5 \times 10^{11}$ Hz
with a parabolic branch.
This latter describes the peak of the SED and its steepening beyond it.
In soft X--rays a rising power law, representing the onset of the hard
inverse Compton component is summed to this curved ``synchrotron" component.
The normalization of this second X--ray component is kept fixed
relative to the radio one. 
Based on our findings (see Fig.~7a), we then assume that the peak
frequency of the synchrotron spectral component is (inversely) related to
radio luminosity.
The simplest hypothesis of a straight unique relationship between 
$\nu_{\rm peak,sync}$ and L$_{\rm 5GHz}$ does not give a good result
when compared with the average SEDs.
We then allow for a different SED--shape/luminosity dependence for 
high and low luminosity objects, a distinction that turns out to roughly
correspond also to that between objects with and without prominent 
emission lines.
We adopted a ``two--branch" relationship between $\nu_{\rm peak,sync}$
and L$_{\rm 5GHz}$ in the form of two power laws $\nu_{\rm peak,sync} 
\propto$ L$_{\rm 5GHz}^{-\eta}$, with $\eta = 0.6$ or $\eta = 1.8$ for 
log(L$_{\rm 5GHz}$) higher or smaller than 42.5, respectively. 
The shape of the analytic SEDs is parabolic with a smooth connection
to a fixed power law in the radio and the loci of the maxima as defined above. 
A full description of the parameterization  can be found in Fossati et al.
(1997) where a similar scheme was proposed to account for the source
number densities of BL Lacs with different spectral properties (LBL and HBL). 

The analytic representation of the second spectral component (X--ray to
$\gamma$--rays) is a parabola of the same width as the synchrotron one,
and has been obtained assuming that:  
(a) the ratio of the frequencies of the high and low energy peaks is constant 
($\nu_{\rm peak, Compt}/\nu_{\rm peak, sync}~\simeq 5 \times 10^8$), 
(b) the high energy ($\gamma$--ray) peak and radio luminosities have a 
fixed ratio, $\nu_\gamma$L$_{\rm peak,gamma} / \nu_{\rm 5GHz}$L$_{\rm 5GHz} 
\simeq 3 \times 10^3$. 
Given the extreme simplicity of the latter assumptions, it is remarkable
that the phenomenological analytic model describes the run of the
average SEDs reasonably well.
The worst case refers to the second luminosity bin: the analytic model
predicts a $\gamma$-ray luminosity larger than the computed bin average by a 
factor of 10 (but predicts the correct spectral shape). We note that 
only 5 $\gamma$--ray detected objects fall in this bin.
 
The results derived from the above analysis (see in particular
Figs.~10--12) can then be summarized as follows: 

\begin{enumerate}
\item {\it two peaks} are present in all the SEDs. The first one
(synchrotron) is anticorrelated with the source luminosity
(see Figs.~7 and Table~4), moving from $\sim 10^{16}-10^{17}$ Hz for
less luminous sources to $\sim 10^{13}-10^{14}$ Hz for the most
luminous ones. 

\item the {\it X--ray spectrum} becomes harder while the $\gamma${\it--ray
spectrum} softens with increasing luminosity, indicating that the second
(Compton) peak of the SEDs also moves to lower frequencies from $\sim
10^{24}-10^{25}$ Hz for less luminous sources to $\sim 10^{21}-10^{22}$ Hz
for the most luminous ones; 

\item therefore {\it the frequencies of the two peaks are correlated}: the
smaller the $\nu_{\rm peak,sync}$ the smaller the peak frequency of the
high energy component; a comparison with the analytic curves shows that
the data are consistent with a constant ratio between the two peak
frequencies;

\item increasing L$_{\rm 5GHz}$ increases the $\gamma${\it--ray
dominance}, \ie the ratio of the power emitted in the inverse Compton and
synchrotron components, estimated with the ratio of their respective  peak
luminosities (see also Fig.~9).  

\end{enumerate}

\noindent

The fact that the trends present when comparing the different samples (\eg
Fig.~10), persist when the total blazar sample is considered and binned
according to radio luminosity only, suggests that we deal with a {\it
continuous spectral sequence} within the blazar family, rather than with
separate spectral classes.  In particular the "continuity" clearly applies
also to the HBL -- LBL subgroups: HBL have the lowest luminosities and the
highest peak frequencies.

An interesting result apparent from  the average SEDs is the variety
and complexity of behaviour shown in the X--ray band.
As expected the crossing between the synchrotron and inverse Compton
components can occur below or above the X-ray band affecting the relation 
between the X--ray luminosity and that in other bands.
A source can be brighter than another at 1 keV being dimmer in the rest of
the radio--$\gamma$--ray spectrum except probably in the TeV range.
This effect narrows the range of values spanned by L$_{\rm 1keV}$
and explains why  $\gamma$--ray detected sources do not select a particular 
range in the X--ray luminosity distributions (see Fig.~3) while this happens 
for L$_{\rm 5GHz}$.

Using this simple scheme of SED parameterization we can compute the 
luminosities in the EGRET (30 MeV -- 3 GeV) and Whipple (0.3 -- 10 TeV) bands. 
These are plotted in Fig.~13 (bottom panel) together with their ratio with
the radio and X--ray luminosities (top and middle panel respectively).

It is easy to recognize that for a given radio flux sources with 
$\nu_{\rm peak,sync}$ around $10^{14}$ have the largest relative 
flux in the EGRET band, because the peak of the Compton component 
falls right there (Fig.~13, top panel). 
For higher values of $\nu_{\rm peak,sync}$ the $\gamma$--ray peak moves to 
higher energies too and the contribution in the EGRET band is reduced. 
For sufficiently high $\nu_{\rm peak,sync}$ the $\gamma$--ray peak reaches 
the TeV band where it becomes detectable. 

Qualitatively the same general behaviour is present also in the
ratios between EGRET/Whipple fluxes and the X--ray one
(Fig.~13, middle panel).  There are however a
couple of significant differences: firstly the EGRET/X--ray ratio
profile, while still peaking around $10^{14}$ Hz, is sharper than in the
EGRET/radio case, secondly the TeV relative flux distribution is
broader and skewed towards lower values of the synchrotron peak frequency.
Thus, for a given X--ray flux (as would be the case in a flux limited
X--ray selected sample) only those sources falling in the restricted
interval $\nu_{\rm peak,sync} \sim 10^{13}-10^{15}$ Hz would have a
flux in the EGRET band high enough to be detectable.
On the other hand, for the TeV band it turns out that
the chance of being observable is not confined to very extreme HBLs,
with X-ray synchrotron peaks, but also intermediate BL Lacs 
could reach a comparable TeV flux.

Since $\nu_{\rm peak,sync}$ is directly related to both $\alpha_{\rm RO}$ 
and $\alpha_{\rm RX}$ we can understand now the tendency (section 3.1) 
of $\gamma$--ray detected sources in the Slew survey to have larger 
values of $\alpha_{\rm RO}$ and $\alpha_{\rm RX}$.
Moreover, due to the fact that in the Slew sample LBLs are only a small 
fraction, the discussion above explains also the lower 
EGRET detection rate with respect 
to other blazar samples.

\begin{figure}
\psfig{file=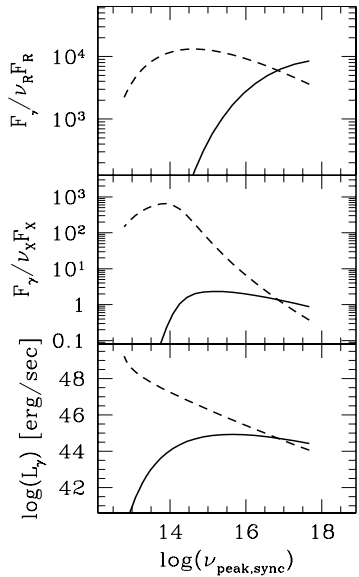,height=8.5truecm,width=9.0truecm,rheight=7.8truecm}
\caption[h] {The predicted relations with $\nu_{\rm peak,sync}$ of the
following quantities: the ratio of the EGRET (dashed line) and Whipple 
(solid line) fluxes (luminosities) with the radio one (top panel); the 
ratio of the EGRET and Whipple fluxes (luminosities) with the X--ray one
(middle panel); EGRET and Whipple absolute luminosities (bottom panel).
The $\gamma$--ray luminosities are integrated in a
band approximately corresponding to that of EGRET or Whipple telescopes,
30 MeV--3 GeV and 0.3--10 TeV, respectively.}
\end{figure}

The proposed scenario relates the shape of the continuum to the total
source power.  It follows that predictions of this unifying scheme on
both the detectability of blazars at $\gamma$--ray energies (in view 
of more sensitive $\gamma$--ray detectors, \eg GLAST, improved Cherenkov 
telescopes, etc.), and their
contribution to the $\gamma$--ray diffuse background depend on the 
combined effects of the SED shape, the luminosity functions and  
possibly evolution (Fossati et al., in preparation; see also 
Stecker, de Jager \& Salamon 1996).  
In particular, an interesting and testable prediction of the scheme
is the absence of high luminosity sources with synchrotron peaks
in the X-ray range and strong associated TeV emission.

\subsection {Interpretation}

The extreme "regularity" of the SEDs of blazars and in particular the
trends discussed above  must derive from the common underlying physical processes.
The common scenario  envisaged is that of a relativistic jet pointing close to the line
of sight. Assuming the simple case of a single (homogeneous) zone model
the shape of the SED depends on the spectrum of the high energy electrons radiating
via synchrotron and inverse Compton, the magnetic field and the nature of seed photons
for the inverse Compton process. The latter could be the synchrotron photons
themselves (synchrotron self Compton, SSC) or photons outside the jet 
(``external Compton" EC). In the following we discuss
qualitatively the implications of the suggested trends for the two scenarios.

Let us assume a constant bulk Lorentz factors in all blazars.  Should
the (homogeneous) SSC model be valid for all sources, it is easy to see that the
(approximately) constant ratio between the high and low peak
frequencies yields an (approximately) constant value for the energy of
the particles radiating at the peaks (e.g. Ghisellini, Maraschi, Dondi
1996).
If the energy of the radiating particles is similar in all sources the
different peak frequencies should result from a systematic variation in
magnetic field strength, HBLs having the highest, FSRQs the lowest random
field intensity.  

Should instead the soft photons upscattered to the $\gamma$--ray range be
produced outside the jet at a "typical" frequency (the same for all
sources) the condition of a constant ratio between the peak frequencies
implies a constant value of the magnetic field (Ghisellini, Maraschi,
Dondi 1996).
As a consequence the energy of the particles radiating at the peaks should
vary along the spectral sequence being lower in FSRQs and higher in HBLs. 

It could also be that there is a smooth transition between the
SSC and EC mechanisms depending on the physical conditions outside the
jet.  In all cases however the role of the luminosity, which
phenomenologically appears dominant, does not find an immediate
physical justification, although one could find plausible arguments to
link it to the parameters mentioned above and in particular to the
conditions surrounding the jet.
 
In a separate paper (Ghisellini et al. 1998), we perform model
fits to the spectral energy distributions of 51 individual objects,
 deriving the (model dependent) physical parameters for each
source.  These computations indeed suggest the idea that the blazar
sequence follows from a transition from the SSC to the EC scenario, RBLs
being the intermediate objects. The computed radiation energy
densities, which determine the amount of radiative cooling, increase
with increasing source luminosity and may be responsible for the lower
energy of the particles radiating at the peaks in higher luminosity
sources.

The likely possibility that the external photon field involved in the
EC process is (or is related to) the radiation reprocessed as broad
emission lines, seems to be at least qualitatively in agreement with
the observational evidence concerning the emission line luminosity in
the suggested blazar sequence.

\section{Conclusions}

The main conclusion of this work is that despite the differences in
the continuum shapes of different sub--classes of blazars, a unitary
scheme is possible, whereby Blazar continua can be described by a
family of analytic curves with the source luminosity as the
fundamental parameter.  The "scheme" (admittedly empirical) 
 determines both the frequency and luminosity of the
peaks in the synchrotron and inverse Compton power distributions
(and therefore also the  $\gamma$--ray luminosity in the EGRET range)
starting from the radio luminosity only. 
The main suggested trend is that with increasing luminosity both
the synchrotron peak and the inverse Compton peak move to lower 
frequencies and that the latter becomes energetically more dominant.
The scheme is testable, for instance it predicts that sources emitting
strongly in the TeV band have relatively low intrinsic luminosity.

The "spectral sequence" finds a plausible interpretation in the framework of
relativistic jet models radiating via the synchrotron and inverse Compton
processes if the physical parameters (magnetic field and/or critical energy
of the radiating electrons) vary with luminosity or if  photons outside the
jet become increasingly important as seed photons for the inverse Compton
process in sources of larger luminosity. The latter alternative is supported
at least qualitatively
by the increasing dominance of emission lines in higher luminosity objects.

The proposed scenario, in which the intrinsic jet power
regulates, in a continuous sequence, the observational properties from
the weaker HBL, through LBL, to the most powerful FSRQs, also fits in
very nicely with the unification of FR~I and FR~II type radio
galaxies as proposed by Bicknell (1995). After a long debate the
prevailing view is that FR~I and FR~II radio galaxies both contain
relativistic jets which can be decelerated giving rise to the FR~I
morphology depending on the kinetic power in the jet and the pressure of
the ambient medium.

The whole radio--loud AGN population could be unified in a two
parameter space one being the intrinsic jet power, the other the
viewing angle. An interesting point for future discussion is whether a
third parameter associated with the luminosity of an accretion disk is
necessary or is already implicitly and uniquely linked to the jet
power.


\section*{Acknowledgments}
The Italian MURST (GF, Annalisa Celotti) and the Institute of
Astronomy PPARC Theory Grant (Annalisa Celotti) are acknowledged for
financial support.  
Andrea Comastri acknowledges financial support from the
Italian Space Agency under contract ARS-96-70
This research has made use of NASA's Astrophysics
Data System Abstract Service and of the NASA/IPAC Extragalactic
Database (NED) which is operated by the Jet Propulsion Laboratory,
Caltech, under contract with the National Aeronautics and Space
Administration.

\section*{References} 

\refitem Adam G., 1985, A\&AS, 1985, 61, 225

\refitem Allen D.A., Ward M.J., Hyland A.R., 1982, MNRAS, 199, 969

\refitem Angel J.R.P., Stockman H.S., 1980, ARA\&A, 8, 321 

\refitem Ballard K.R., Mead A.R.G., Brand P.W.J.L., Hough J.H., 1990, MNRAS, 243, 640

\refitem Becker R.L., White R.L., Edwards A.L., 1991, ApJS, 75, 1

\refitem Bersanelli, M., Bouchet P., Falomo R., Tanzi E.G., 1992, AJ, 104, 28

\refitem Bertsch D.L. et al., 1993, ApJ, 405, L21

\refitem Bicknell G.V., 1995, ApJS, 101, 29

\refitem Blandford R.D., 1993, in Friedlander M., Gehrels N., Macomb D.J., eds, Proc. CGRO AIP 280. New York, p. 533

\refitem Bloom S.D., Marscher A.P., 1993, in Friedlander M., Gehrels N., Macomb D.J., eds, Proc. CGRO AIP 280. New York, p. 578

\refitem Bloom S.D., Marscher A.P., Gear W.K., Terasranta H., Valtaoja E., Aller H.D., Aller M.F. 1994, AJ, 108, 398

\refitem Bradbury S.M. et al., 1997, A\&A, 320, L5

\refitem Brindle C., Hough J.H., Bailey J.A., Axon D.J., Hyland A.R., 1986, MNRAS, 221, 739

\refitem Brinkmann W., Siebert J., Boller T., 1994, A\&A, 281, 355

\refitem Brinkmann W. et al., 1995, A\&AS, 109, 147

\refitem Brown L.M.J. et al., 1989, ApJ, 340, 129

\refitem Brunner H., Lamer G., Worrall D.M., Staubert R., 1994, A\&A, 287, 436 

\refitem Catanese M. et al., 1997, ApJ, 480, 562

\refitem Chiang J, Fichtel C.E., von Montigny C., Nolan P.L., Petrosian V.,1995, ApJ, 452, 156

\refitem Comastri A., Molendi S., Ghisellini G., 1995, MNRAS, 277, 297 

\refitem Comastri A., Fossati G., Ghisellini G., Molendi S., 1997, ApJ, 480, 534

\refitem Dermer C.D., Schlickeiser R., 1993, ApJ, 416, 458

\refitem Dickey J. M., Lockman F. J., 1990, ARA\&A, 28, 215

\refitem Dingus B.L. et al., 1996, ApJ, 467, 589

\refitem Dondi L., Ghisellini, G. 1995, MNRAS, 273, 583

\refitem Elvis M., Lockman F. J., Wilkes B. J. 1989, AJ, 97, 777

\refitem Elvis M., Plummer D., Schachter J., Fabbiano G., 1992, ApJS,
80, 257

\refitem Elvis M. et al., 1994, ApJS, 95, 1

\refitem Falomo R., Bersanelli M., Bouchet P., Tanzi E.G., 1993a, AJ, 106, 11

\refitem Falomo R., Treves A., Chiappetti L., Maraschi L., Pian E., Tanzi E.G., 1993b, ApJ, 402, 532L

\refitem Falomo R., Scarpa R., Bersanelli M., 1994, ApJS, 93, 125

\refitem Fegan D.J., 1996, private communication 

\refitem Fichtel C.E. et al., 1994, ApJS, 94, 551

\refitem Fossati G., Celotti A., Ghisellini G., Maraschi L., 1997, MNRAS, 289, 136 

\refitem Gabuzda D.C., Cawthorne T.V., Roberts D.H., Wardle J.F.C.,
1992, ApJ, 338, 40

\refitem Gear W.K., 1993a, MNRAS, 264, L21

\refitem Gear W.K., 1993b, MNRAS, 264, 919

\refitem Gear W.K. et al., 1985, ApJ, 291, 511

\refitem Gear W.K. et al., 1986, ApJ, 304, 295 

\refitem Gear W.K. et al., 1994, MNRAS, 267, 167

\refitem Ghisellini G., Madau P., Persic M., 1987, MNRAS, 224, 257

\refitem Ghisellini G., Madau P. 1996, MNRAS, 280, 67

\refitem Ghisellini G., Maraschi L., Dondi L., 1996, A\&AS, 120, 503

\refitem Ghisellini G., Padovani P., Celotti A., Maraschi L., 1993,
ApJ, 407, 65

\refitem Ghisellini G., Celotti A., Fossati  G., Maraschi L.,  Comastri A.,
1998, MNRAS, submitted 

\refitem Giommi P., Padovani P., 1994, MNRAS, 268, L51

\refitem Glass I.S., 1979, MNRAS, 186, L29

\refitem Glass I.S., 1981, MNRAS, 194, 795

\refitem Hartman R.C. et al., 1993, ApJ, 407, L41

\refitem Holmes P.A., Brand P.W.J.L., Impey C.D., Williams P.M., 1984, MNRAS, 210, 961

\refitem Impey C.D. 1996, AJ, 112, 2667

\refitem Impey C.D., Brand P.W.J.L., 1981, Nature, 292, 814

\refitem Impey C.D., Brand P.W.J.L., 1982, MNRAS, 201, 849

\refitem Impey C.D., Brand P.W.J.L., Wolstencroft R.D., Williams P.M., 1982, MNRAS, 200, 19

\refitem Impey C.D., Brand P.W.J.L., Wolstencroft R.D., Williams P.M., 1984, MNRAS, 209, 245

\refitem Impey C.D., Neugebauer G., 1988, AJ, 95, 307 

\refitem Impey C.D., Tapia S., 1988, ApJ, 333, 666

\refitem Impey C.D., Tapia S., 1990, ApJ, 354, 124

\refitem Jannuzi B.T., Smith P.S., Elston R., 1993, ApJS, 85, 265

\refitem Jannuzi B.T., Smith P.S., Elston R., 1994, ApJ, 428, 130

\refitem Jones T.W., O'Dell S.L., Stein W.A., 1974, ApJ, 188, 353

\refitem K\"uhr H., Witzel A., Pauliny--Toth I.K., Nauber U., 1981,
A\&A, 45, 367

\refitem Lamer G., Brunner H., Staubert R., 1996, A\&A, 311, 384

\refitem Landau R. et al., 1986, ApJ, 308, 78

\refitem Lepine J.R.D., Braz M.A., Epchtein N., 1985, A\&A, 149, 351

\refitem Lin Y.C. et al., 1995, ApJ, 442, 96 

\refitem Lichtfield S.J., Robson E.I., Stevens J.A., 1994, MNRAS, 270, 341

\refitem Lockman F. J., Savage B. D. 1995, ApJS, 97,1

\refitem Lorenzetti D., Massaro E., Perola G.C., Spinoglio L., 1990, A\&A, 235, 35

\refitem Madejski G, et al., 1996, ApJ, 459, 156

\refitem Maraschi L., Rovetti F., 1994, ApJ, 436, 79

\refitem Maraschi L., Fossati G., Tagliaferri G., Treves A., 1995, ApJ, 443, 578

\refitem Maraschi L., Ghisellini G., Celotti A., 1992, ApJ, 397, L5

\refitem Maraschi L., Ghisellini G., Tanzi E.G., Treves A., 1986, ApJ, 310, 325

\refitem Mattox J.R. et al., 1997, ApJ, 476, 692

\refitem Mattox J.R., Schachter J., Molnar L., Hartman R.C., Patnaik
A.R., 1997, ApJ, 481, 95

\refitem Mead A.R.G., Ballard K.R., Brand P.W.J.L., Hough J.H., Brindle C., Bailey J.A., 1990, A\&AS, 83, 183

\refitem Morris S.L., Stocke J.T., Gioia I.M., Schild R.E., Wolter A.,
Maccacaro T., Della Ceca R., 1991, ApJ, 380, 49

\refitem M\"ucke A. et al., 1997, A\&A, 320, 33

\refitem Mukherjee R. et al., 1995, ApJ, 445, 189

\refitem Mukherjee R. et al., 1996, ApJ, 470, 831

\refitem Murphy E.M., Lockman F.J., Laor A., \& Elvis M. 1996, ApJS, 105, 369

\refitem Nolan P.L. et al., 1993, ApJ, 414, 82  

\refitem Nolan P.L. et al., 1996, ApJ, 459, 100 

\refitem O'Dell S.L., Puschell J.J., Stein W.A., Warner J.W., 1978, ApJS, 38, 267

\refitem Padovani P., 1992a, MNRAS, 257, 404

\refitem Padovani P., 1992b, A\&A, 256, 399

\refitem Padovani P., Giommi P., 1995, ApJ, 444, 567

\refitem Padovani P., Urry C.M., 1992, ApJ, 387, 449 

\refitem Pian E., Falomo R., Scarpa R., Treves A., 1994, ApJ, 432, 547

\refitem Perlman E.S., Stocke J.T., Schachter J.F., Elvis M., Ellingson E., Urry C.M., Potter M., Impey C.D., Kolchinsky P., 1996a, ApJS, 104, 251   (SLEW)

\refitem Perlman E.S., Stocke J.T., Wang Q.D., Morris S.L., 1996b, ApJ, 456, 451 (EMSS)

\refitem Quinn J. et al., 1996, ApJ, 456, L83

\refitem Radecke H.--D. et al., 1995, ApJ, 438, 659

\refitem Reuter H.-P. et al., 1997, A\&AS, 122, 271

\refitem Riecke G.H., Lebofski M.J., 1985, ApJ 288, 618

\refitem Sambruna R.M., 1997, ApJ, 487, 536

\refitem Sambruna R.M., Maraschi L., Urry C.M., 1996, ApJ, 463, 444

\refitem Scarpa , Falomo R., 1997, A\&A, 325, 109

\refitem Shrader C.R., Hartman R.C., Webb J.R., 1996, A\&AS, 120, 599

\refitem Sikora M., Begelman M.C., Rees M.J., 1994, ApJ, 421, 153

\refitem Sitko M.L., Schmidt G.D., Stein W.A., 1985, ApJS, 59, 323

\refitem Sitko M.L., Sitko A.K., 1991, PASP, 103, 160

\refitem Smith P., Balonek T.J., Elston R., Heckert P.A., 1987, ApJS, 64, 459

\refitem Sreekumar P. et al., 1996, ApJ, 464, 628

\refitem Stecker F.W., de Jager O.C., Salamon M.H., 1996, ApJ, 473, L75

\refitem Steppe H. et al., 1988, A\&AS,  75, 317

\refitem Steppe H. et al., 1992, A\&AS,  96, 441

\refitem Steppe H. et al., 1993, A\&AS, 102, 611

\refitem Stevens J.A. et al 1994,  ApJ, 437, 91

\refitem Stickel M. et al., 1991, ApJ, 374, 431

\refitem Stickel M., Meisenheimer, K., K\"uhr H.  1994, A\&AS, 105, 211

\refitem Terasranta H. et al., 1992, A\&AS, 94, 121

\refitem Thompson D.J. et al., 1993, ApJ, 415, L13

\refitem Thompson D.J. et al., 1995, ApJS, 101, 259

\refitem Thompson D.J. et al., 1996, ApJS, 107, 227

\refitem Tornikovski M. et al., 1993, AJ, 105, 1680

\refitem Tornikovski M. et al., 1996, A\&AS, 116, 157

\refitem Ulrich M.--H., Maraschi L., Urry C.M., 1997, ARA\&A, 35

\refitem Urry C.M., Padovani P., 1995, PASP, 107, 803

\refitem Urry C.M., Sambruna R.M., Worrall D.M., Kollgaard R.I., Feigelson E.D., Perlman E.S., Stocke J.T., 1996, ApJ, 463, 424

\refitem Vestrand, W.T., Stacy J.G., Sreekumar P., 1995, ApJ, 454, L93

\refitem von Montigny C. et al., 1995, ApJ, 440, 525

\refitem Wall J.V., Peacock J.A., 1985, MNRAS, 216, 173

\refitem Weekes T.C. et al., 1996, A\&AS, 120, 603

\refitem Wolter A., Caccianiga A., Della Ceca R., Maccacaro T., 1994,
ApJ, 433, 29

\refitem Worrall D.M., Wilkes B.J., 1990, ApJ, 360, 396

\refitem Wright A., Ables J.G., Allen D.A., 1983, MNRAS, 205, 793

\vfill\eject

\end{document}